\documentclass[aps,pre,superscriptaddress]{revtex4-1}
\usepackage{dcolumn}% Align table columns on decimal point
\usepackage{bm}% bold math
\usepackage{textcomp}
\usepackage{amsmath}
\usepackage{graphicx}
\usepackage{subfigure}
\usepackage{color}
\usepackage[latin1]{inputenc}
\usepackage[T1]{fontenc}
\usepackage{lmodern}
\usepackage[rgb]{xcolor}
\usepackage{pdfcomment}

\def\r1{\hat{\rho}_1}
\begin{document}

\title{A semi-flexible model prediction for the polymerization force exerted by a living F-actin filament on a fixed wall.}
\author{Carlo Pierleoni}
\email{carlo.pierleoni@aquila.infn.it}
\affiliation{Department of Physical and Chemical Sciences, University of L'Aquila, and CNISM UdR L'Aquila, Via Vetoio 10, 67100 L'Aquila, Italy}

\author{Giovanni Ciccotti}
\email{giovanni.ciccotti@roma1.infn.it}
\affiliation{Physics Department, Sapienza University of Rome, P. A. Moro 5, 00185 Rome, Italy}
\affiliation{School of Physics, University College Dublin (UCD), Belfield, Dublin 4, Ireland}
\author{Jean-Paul Ryckaert}
\email{jryckaer@ulb.ac.be}
\affiliation{Physics Department, Universit\'e Libre de Bruxelles (ULB), Campus Plaine, CP 223, B-1050 Brussels, Belgium}

\date{\today}
%  but any date may be explicitly specified

%\contributor{Submitted to Proceedings of the National Academy of Sciences of the United States of America}

%%%Newly updated.
%%% If significance statement need, then can use the below command otherwise just delete it.
%\significancetext{CP, GC and JPR bla bla bla.}

%\begin{article}
%%%%%%%%%%%%%%%%%%%%%%%%%%%%%%%%%%%%%%%%%%%%%%%%%%%%%%%%%%%%%%%%%%%%%%%%%%%
\begin{abstract}
We consider a single living semi-flexible filament with persistence length $\ell_p$ in chemical equilibrium with a solution of free monomers at fixed monomer chemical potential $\mu_1$ and fixed temperature $T$. While one end of the filament is chemically active with single monomer (de)polymerization steps, the other end is grafted normally to a rigid wall to mimick a rigid network from which the filament under consideration emerges. A second rigid wall, parallel to the grafting wall, is fixed at distance $L<<\ell_p$ from the filament seed. In supercritical conditions where monomer density $\rho_1$ is higher than the critical density $\rho_{1c}$, the filament tends to polymerize and impinges onto the second surface which, in suitable conditions (non-escaping filament regime) stops the filament growth. We first establish the grand-potential $\Omega(\mu_1,T,L)$ of this system treated as an ideal reactive mixture and derive some general properties, in particular the filament size distribution and the force exerted by the living filament on the obstacle wall. We apply this formalism to the semi-flexible, \textit{living}, discrete Wormlike chain (d-WLC) model with step size $d$ and persistence length $\ell_p$, hitting a hard wall. Explicit properties require the computation of the mean force $\bar{f}_i(L)$ exerted by the wall at $L$ and associated potential $\bar{f}_i(L)=-dW_i(L)/dL$ on a filament of fixed size $i$. By original Monte-Carlo calculations for few filament lengths in a wide range of compression, we justify the use of the weak bending universal expressions of Gholami et al.(Phys.Rev.E. 74,(2006), 041803) over the whole non escaping filament regime. For a filament of size $i$ with contour length $L_c=(i-1)d$, this universal form is rapidly growing from zero (non compression state) to the buckling value $f_b(L_c,\ell_p)=\frac{\pi^2 k_BT \ell_p}{4 L_c^2}$ over a compression range much narrower than the size $d$ of a monomer. Employing this universal form for living filaments, we find that the average force exerted by a living filament on a wall at distance $L$ is in practice $L$ independent and very close to the value of the stalling force $F_s^H=(k_BT/d) \ln(\hat{\rho}_1)$ predicted by Hill, this expression being strictly valid in the rigid filament limit. The average filament force results from the product of the cumulative size fraction $x=x(L,\ell_p,\hat{\rho}_1)$, where the filament is in contact with the wall, times the buckling force on a filament of size $L_c \approx L$, namely $F_s^H=x f_b(L;\ell_p)$. The observed $L$ independence of $F_s^H$ implies that $x \propto L^{-2}$ for given ($\ell_p,\hat{\rho}_1$) and $x \propto \ln{\hat{\rho}_1}$ for given ($\ell_p,L$). At fixed ($L,\hat{\rho}_1$), one also has $x \propto \ell_p^{-1}$ which indicates that the rigid filament limit $\ell_p \rightarrow\infty$ is a singular limit in which an infinite force has zero weight. Finally we derive the physically relevant threshold for filament escaping in the case of actin filaments.

\end{abstract}
%%%%%%%%%%%%%%%%%%%%%%%%%%%%%%%%%%%%%%%%%%%%%%%%%%%%%%%%%%%%%%%%%%%%%%%%%%%
%\keywords{monolayer | structure | x-ray reflectivity | molecular electronics}

%\abbreviations{SAM, self-assembled monolayer; OTS, octadecyltrichlorosilane}

%\pacs{Valid PACS appear here}% PACS, the Physics and Astronomy
                             % Classification Scheme.
%\keywords{Suggested keywords}%Use showkeys class option if keyword
                              %display desired
%\tableofcontents

\maketitle

%
%  SECTION INTRO
%
%
\section{Introduction. \label{sec:intro}}

Cytoskeleton actin filaments, with the help of a wide variety of auxiliary proteins, are at the root of dynamical processes involved in cell motility\cite{b.Howard,SZ.05}. The growth of lamellipodium and filopedia is directly related to actin filaments pushing or sometimes pulling (with the help of trans membrane proteins) with their barbed end pointing against the cellular membrane. A subtle interplay of polymerizing or depolymerizing steps, involving single G-actin monomers at the barbed end, provides the essential mechanism allowing the cytoskeletal network to keep contact while maintaining a permanent pressure force on a load resisting membrane.

In vitro experiments on biofilaments, like actin and tubulin, to measure in supercritical conditions either the force-velocity relationship in detectable non-zero velocity conditions\cite{Dogterom.97,DCBB.14}, or the approach to stalling where the applied load effectively stops the net polymerization of living filaments and the stalling force is effectively measured\cite{Dogterom.07}, have been deviced. To simplify the analysis and to concentrate on the fundamental process of force generation by polymerizing filaments, the experiments deal with bundles of parallel filaments hitting an orthogonal moving wall, a network having strong analogy with the structure of actin filopedia\cite{b.Howard}. Data analysis requires models for bundle dynamics and stalling force predictions, and in general, most models treat living filaments as perfectly rigid.

In a series of pioneering papers on this topics in the eighties, Hill was the first to propose the expression~\cite{Hill.81}
\begin{align}
F_{bun}^H=N_f \frac{k_BT}{d} \ln{\hat{\rho}_1}
\label{eq:force}
\end{align}
for the force generated by a bundle of $N_f$ growing (proto)filaments stopped by a normal wall. In Eq.(\ref{eq:force}), $T$ is the absolute temperature, $d$ is the effective monomer size along the filament contour (equal to half the G-actin diameter as there are two interwined protofilaments in F-actin) and $\hat{\rho}_1=U_0/W_0>1$ is the reduced free monomer density equal to the ratio of bulk polymerizing and depolymerizing rates $U_0$ and $W_0$. Using a combination of thermodynamic and mean-field arguments \cite{Hill.81}, this expression has been established as the equilibrium state (zero growth velocity) of a more general expression linking the wall velocity and the load force for a bundle of filaments in a generally non equilibrium framework. As stressed by Hill, eq. (\ref{eq:force}) is derived from a one-dimensional longitudinal and incompressible model which implies a proportionality between the average polymerization rate and the average wall velocity. 

The 1D brownian ratchet model for an individual rigid filament hitting a moving wall was later proposed to offer a physically justified stochastic model~\cite{POO.93}. The living filament is subject to random polymerizing and depolymerizing steps with respective rates $U_0$ and $W_0$ with $U_0>W_0$ to treat supercritical conditions. While the depolymerizing step is possible even in presence of the wall, the polymerizing step is only accepted if it does not lead to an overlap with the moving wall. In addition the wall undergoes a 1D brownian motion characterized by a diffusion coefficient and by a load which biases the wall dynamics towards the filament's end. The coupling between the filament (de)polymerization dynamics and the wall random motion leads to a stalling force in agreement with Eq.(\ref{eq:force}) and to a stationary drift velocity of the wall, which for large wall diffusion coefficient, agrees with Hill's prediction of the load-velocity law. When many parallel filaments act together as a bundle, the brownian rachet model can be generalized to a multi-rigid filament system while remaining essentially 1D \cite{MO.99,sander.00,Joanny.11,DCBB.14}. The dynamical coupling among filaments via the common wall evolution, which is very sensitive to the relative longitudinal disposition of the filaments, has strong implication on the velocity-load relationship \cite{Dogterom.97,sander.00,DCBB.14}. Let us note that all the above models consider non interacting filaments and a single kind of actin-Adenosine triphosphate (actin-ATP) complex for the monomers whether free or incorporated into filaments. The above dynamical models can be generalized to take into account lateral interactions among filaments \cite{Krawczyk.11,LK.15} mainly to treat a many-protofilaments model and/or the hydrolysis of the ATP(Guanosine triphosphate-GTP) in the filament actin-complexes (tubulin-complexes) by considering additional types of complexes, requiring in turn additional information on specific rates \cite{Ca.01,DDP.14,LK.15}.

If rigid filament models are certainly satisfactory as long as the elementary working filaments (being isolated and uncrosslinked) remain sufficiently short, the flexibility of F-actin should be properly considered for longer filaments. Flexibility was found to be relevant in some important experimental cases. 
%\pdfmarkupcomment[markup=StrikeOut,color=red]{and in some theoretical predictions. We first note that}{}
The bending shape of single F-actin filaments observed by fluorescence spectroscopy, was precisely exploited to measure for the first time the typical polymerization force generated by single living actin filaments \cite{KP.04,Berrot.07}. In the optical trap experiment of Footer et al.\cite{Dogterom.07}, a bundle of about ten filaments, with their seeds glued to a trapped colloidal particle, push with their active side (barbed end) on a fixed rigid wall. The polymerization force they progressively develop to reach equilibrium is inferred by measuring the colloidal displacement in the trap. 
%\pdfmarkupcomment[markup=StrikeOut,color=red]{The production of data and their subsequent analysis were made very difficult as a result of the detected presence of escaping filaments (filaments growing parallel to the obstacle wall after a large angle bending fluctuation), a phenomenon interpreted by the authors as rod buckling related the beam elasticity instability}{}
The interpretation of the experiments was possible only by assuming the presence of escaping filaments in the bundle (filaments growing parallel to the obstacle wall after a large angle bending fluctuation), a phenomenon interpreted by the authors as rod buckling related the beam elastic instability
\cite{b.Howard}. Despite care in eliminating data potentially polluted by escaping filaments, the measured stalling force for a eight filament bundle was (repeatedly) found to be close to Eq.(\ref{eq:force}) with $N_f\approx 1$ instead of the expected $N_f=8$ filaments number, a result still presently not understood.

That flexibility leads, in some extreme cases, to escaping filaments was reported and analyzed in a non equilibrium simulation of a model of single living filament hitting a moving wall in which filament flexibility was explicitly taken into account\cite{BM.05,BM.06}. 
Quite generally, in these pseudo-stationary simulations with constant load, a wall velocity enhancement was found with respect to the predictions of the "rigid filament-hard wall" ratchet model, in agreement with theoretical considerations which have generally predicted an enhancement of the efficiency of the conversion of chemical free energy into useful work when realistic filament flexibility is included\cite{MO.96,SB.08}. For large loads (still below the stalling force) and for large seed-wall distances, some escaping filaments were detected during the drift of the wall \cite{BM.05,BM.06}. It was argued that this phenomenon is related but distinct from rod buckling and hence was denoted as the "pushing catastrophe". The consensus seems to be that to efficiently grow against membrane resistance, actin filaments should be neither too short (short filaments are too rigid to intercalate easily a polymerizing monomer between the tip of the filament and the wall) nor too long as the load would simply buckle them, the optimal range $70nm-500nm$ being cited in a recent review article\cite{Mo.09}.

In this paper we concentrate on the equilibrium Statistical Mechanical treatment of a semi-flexible filament in a slab.
%\pdfmarkupcomment[markup=StrikeOut,color=red]{We present a Statistical Mechanics analysis of a single grafted living semi-flexible filament hitting a fixed wall along lines initiated by one of us in a previous study, limited however to some general features of living filaments in the canonical ensemble.}{}
In section \ref{sec:therm}, extending previous work \cite{RRMP.13}, we establish within the reactive grand canonical ensemble, the grand potential for a living filament in contact with an obstacle wall at fixed temperature and fixed free monomer chemical potential. 
%\pdfmarkupcomment[markup=StrikeOut,color=red]{In the non-escaping equilibrium conditions, the partition function of a single grafted living filament within a slab of width $L$ comprises contributions from configurations of long filaments in contact with the obstacle wall and configurations of short filaments which do not interact with the obstacle.}{}
Formal expressions for the size distribution and the equilibrium force on the obstacle wall are established. 
Section \ref{actin} deals specifically with F-actin modeled as a living discrete Wormlike Chain (d-WLC). We first define the model and the related range of physical parameters to probe the non-escaping regime of the filament. We then compute, by Monte Carlo simulation, the compression-force law for a dead (non-reacting) d-WLC in the slab and validate, in the non-escaping regime, the weak bending expression of Gholami et al \cite{Frey.06}.
%\pdfmarkupcomment[markup=StrikeOut,color=red]{which will be used throughout the remaining part of the study.}{} 
Subsequently we define the filament force averaged over a distance equal to a monomer size $d$, crucial for the comparison with Hill's prediction, 
%\pdfmarkupcomment[markup=StrikeOut,color=red]{and we define quantitatively}{} together with the limit of the non-escaping regime for F-actin at given supercritical conditions. 
In section \ref{comparison} we introduce the stalling force and compare the predictions for flexible (finite $\ell_p$) against rigid ($\ell_p\to \infty$) models, proving for the latter Hill's expression for the stalling force.
In the entire range of filament lengths corresponding to the non-escaping regime, the flexible filament has a stalling force only few percents larger than a rigid filament (Hill's law). Nonetheless, the specific $L$-dependence of the force ($\sim L^{-2}$) resulting from buckled filaments hitting the obstacle wall, induces a spectacular, previously undetected, effect of flexibility. Since the stalling force is nearly independent of the slab's width $L$ in the non-escaping regime, this requires a systematic evolution with $L^2$ of the fraction of sizes of the filament touching the wall. This is discussed in section \ref{distr}.
%\pdfmarkupcomment[markup=StrikeOut,color=red]{for which the standard Hill's result for the stalling force is demonstrated within the present equilibrium Statistical Mechanics framework. In section ..., a closer look to the $L$ dependence of the filament size distribution, together with considerations on filament buckling, allow us to provide an unexpected consequence of filament flexibility, namely a systematic evolution with $L$ of the fraction of sizes of the filament touching the wall.}{}
Finally, section \ref{discussion} provides some general conclusions and perspectives on the flexibility issue for many filaments bundles, including both static properties 
%\pdfmarkupcomment[markup=StrikeOut,color=red]{similar to those covered here for a single filament, but also}{} 
and dynamic aspects linked directly to the exploration of the force-load relationship.

%
%SECTION 2
%
%%%%%%%%%%%%%%%%%%%%%%%%%%%%%%%%%%%%%%%%%%%%%%%%%%%%%%%%%%%%%%%%%%%%%%%%%%%
\section{The single grafted living filament in a slab system \label{sec:therm}}
%%%%%%%%%%%%%%%%%%%%%%%%%%%%%%%%%%%%%%%%%%%%%%%%%%%%%%%%%%%%%%%%%%%%%%%%%%%

\subsection{The single grafted living filament concept}

We consider a reacting ideal mixture in a slab at temperature $T$ consisting of $N_t$ monomers which can either be free (G actin-ATP complex) or integrated within a single self-assembled filament (F-actin) with fixed persistence length $\ell_p$. In the F-actin case, $\ell_p=5370d$ and $d=2.7nm$ is the effective monomer size in the filament. The filament, with a variable size $i$ and associated contour length $L_{\rm c}=(i-1)d$, is grafted normally to one of the walls of the slab considered as an orthorhombic volume of transverse area $A$ and width (wall to wall distance) $L<<\ell_p$. 
The filament undergoes single monomer (de)polymerization events with a polymerization rate $U_0 = k_{on}\rho_1$, proportional to the free monomers density $\rho_1$, and a depolymerization rate $W_0 = k_{off}$, independent on the free monomer density, where $k_{on/off}$ are the kinetic constants for the (de)polymerization reactions. Supercritical conditions are realized whenever the bulk polymerization rate is larger than the depolymerization rate, which happens for $\rho_1 > k_{off} /k_{on} = 1/K_0$, where $K_0$ is the bulk reaction equilibrium constant\cite{b.Hill}. We define $\hat\rho_1\equiv\rho_1K_0$ as the reduced free monomers density; supercritical conditions correspond to $\hat\rho_1>1$.
In super-critical conditions where polymerization dominates, the filament will grow and hit the opposite wall as soon as $L_{\rm c}>L$.
\begin{figure}[ht]
\begin{center}
\includegraphics[angle=0,scale=0.90]{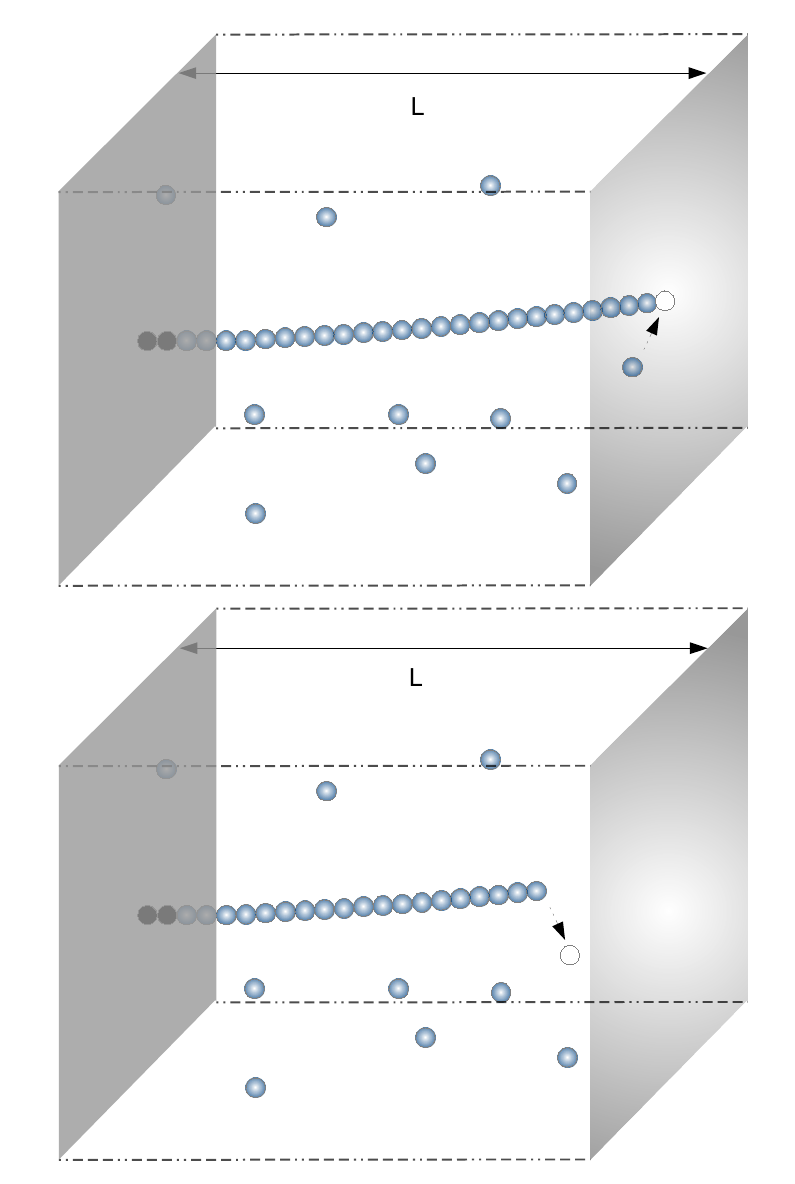}\\
\caption{Two configurations of the same living filament grafted normally to the left wall through its first two monomers drawn in dark grey, in chemical equilibrium with a free monomers solution. According to the whole set of (de)polymerizing reactions Eqs.(\ref{eq:reaction}), the filament can polymerize or depolymerize at its free end by addition or removal of one single monomer as illustrated by the arrows. In our illustrations (a,b), the contour length $L_c$ of the polymerizing filament in (a) appears to be longer than the distance $L$ between the walls while in (b), the contour length $L_c$ of the depolymerizing filament is shorter than the distance $L$.}
\label{filament}
\end{center}
\end{figure}

The series of possible chemical reaction will be denoted as
\begin{align}
A_{i-1} + A_1 \rightleftharpoons A_{i}      \ \ \ (2<i \leq z^{*})
\label{eq:reaction}
\end{align}
where $A_i$ and $A_1$ represent respectively the grafted filament of size $i$ 
and a free monomer. At global equilibrium, the chemical potentials $\mu_i$ of the different species involved in any reaction must satisfy the chemical equilibrium requirement
\begin{align}
\mu_{i}=\mu_{i-1}+\mu_1
\label{eq:chemeq}
\end{align}
This series of reactions is considered as limited to a size window going from 
a minimum filament size of two (to be considered as an effective permanent 
seed of the filament) up to a maximum size $z^*$. Fixing a maximum filament length is only necessary when considering flexible filaments. In fact for rigid 1D filaments, the obstacle hard wall will necessarily limit the filament growth. Instead for flexible filaments, equilibrium Statistical Mechanics based on the concept of a steady equilibrium state, can only be applied if we consider a mechanism limiting the filament growth in particular if supercritical conditions are considered. Again if the slab is narrow enough with respect to the filament persistence length, the obstacle wall will effectively limit the filament growth but for wider slabs we need to introduce an artificial limit. In our present geometry (see fig. \ref{filament}) we impose a maximum filament size
\begin{equation}
z^{*}=INT\left(\frac{\pi L}{2 d}\right).
\label{eq:z*}
\end{equation}
where $INT(x)$ means the integer part of the argument. 
%\pdfmarkupcomment[markup=StrikeOut,color=red]{Introducing a maximum filament size is necessary because}{} 
%Filaments longer than $z^{*}$, \CP{the size corresponding to a planar filament} configuration which covers a quarter of a circle of radius $L$, 
In absence of this size limit, corresponding to filaments adopting a planar configuration which covers a quarter of a cycle of radius $L$, filaments longer than z* could grow
%\pdfmarkupcomment[markup=StrikeOut,color=red]{in supercritical conditions can grow without limit}{} 
unhindered parallel to obstacle wall, preventing the establishment of an equilibrium state. 
%\pdfmarkupcomment[markup=StrikeOut,color=red]{On the other hand, t}{} 
The imposed upper limit will bias the statistical mechanics averages except for suitable external conditions (choice of control variables $N_t/AL$ and $L$ in particular) for which the statistical weight of filaments of size $z^{*}$ or longer is negligibly small. This defines what we call the "non-escaping" regime at stalling conditions.

In terms of the temperature $T$, volume $V=AL$, total number of monomers $N_t$ and total number of grafted filaments $N_f$, the reversible change of the relevant Helmholtz thermodynamic potential $F^{R}$ for this reactive system (hence the R superscript) is
\begin{equation}
dF^{R}=-S dT - p_N A dL + p_T L dA + \mu_1 dN_t + (\mu_2-2 \mu_1)dN_f
\label{eq:F}
\end{equation}
where $S$ is the system entropy and where $p_N A$ and $p_T A$ are the total normal and tangential forces exerted by the obstacle wall on the system. The last two terms involve the chemical potential $\mu_1$ of free monomers and the chemical potential $\mu_2$ of grafted filaments (seeds) of minimum size $2$. These two last terms result from imposing chemical equilibrium Eqs.(\ref{eq:chemeq}) for all reactions (\ref{eq:reaction}) to the original series of terms $\sum_{i=1}^{z^{*}} \mu_i dN_i$ involving all species of the mixture. In the latter sum, the chemical potential of any filament with size $i>2$ is substituted by $\mu_i=\mu_2+(i-2)\mu_1$ which follows recursively from Eqs.(\ref{eq:chemeq}) and a final regrouping of terms yields Eq.(\ref{eq:F}), taking into account the expressions of the total number of monomers $N_t=N_1+\sum_{i=2}^{z^{*}}\;i \;N_i$ and of the total number of filaments $N_f=\sum_{i=2}^{z^{*}}N_i$.

\subsection{Free energy of a grafted living filament under confinement}

Applying equilibrium statistical mechanics to a closed reacting ideal system\cite{b.Hill}, the canonical partition function $Q^{\rm R}=\exp{[-\beta F^{R}]}$ for a single grafted filament in a solution of free monomers is given by 
\begin{align}
Q^{\rm R}(A,L,T,N_t,N_f=1)&= \frac{q_1^{(N_t-2)}}{(N_t-2)!} q_2 + \frac{q_1^{(N_t-3)}}{(N_t-3)!} q_3+..
\frac{q_1^{(N_t-z^{*})}} {(N_t-z^{*})!} q_{z^{*}}
\label{eq:rci1}
\end{align}
The sum over all distinct microscopic states compatible with the macroscopic variables is expressed in Eq.(\ref{eq:rci1}) as a sum over $(z^{*}-1)$ similar terms, each of them corresponding to one particular size of the single grafted filament and the remaining free monomers. Each term of this ideal system involves the canonical partition function $q_i(L,T)$ of the filament of size $2 \leq i \leq z^{*}$ grafted in the slab and the corresponding contribution from the free monomers
\begin{align}
q_1(L,A,T)=\frac{AL}{\Lambda^3} 
\label{eq:q1}
\end{align}
where $\Lambda$ is the free monomer thermal de Broglie wavelength. 

To each term $q_i(L,T)$ corresponds a canonical partition functions $q_{i}^{0}(T)$ relative to the same grafted filament of size $i$ in the absence of the opposite wall. Keeping the temperature dependence implicit, we define the ratio's 
\begin{align}
\alpha(i,L)=\frac{q_i(L)}{q_{i}^0}
\label{eq:alpha}
\end{align}
As long as the intra-filament interactions have a local and homogeneous character, the ratio between successive partition functions $q_{i-1}^0$ and $q_i^0$ is independent of $i$. Hence, as further detailed in section \ref{reaction}, we introduce a temperature dependent equilibrium constant $K_0$ \cite{b.Hill}
\begin{align}
K_0& \equiv \frac{q_i^0}{q_{i-1}^0 q_1/V} =\frac{q_i^0}{q_{i-1}^0}\Lambda^3
\label{eq:K0}
\end{align}
Using Eqs.(\ref{eq:alpha},\ref{eq:K0}), the partition functions of the filaments of any size $i$ can be written as
\begin{align}
q_{i}(L)&= \alpha(i,L) q_2^0 
\left(\frac{K_0}{\Lambda^3}\right)^{i-2} \ (2 \leq i \leq z^{*}).
\label{eq:qif1} 
\end{align}
where $q_2^0$ is the partition function of the grafted seed.

Eq.(\ref{eq:rci1}) can now be combined with expressions (\ref{eq:q1},\ref{eq:qif1}), to give
\begin{align}
Q^{\rm R}(A,L,T,N_t,N_f=1)&=q_2^0 q_1^{N_t-2} \times \left[ \frac{\alpha(2,L)}{(N_t-2)!} + \frac{q_1^{-1} \alpha(3,L)\left(\frac{K_0}{\Lambda^3}\right)}{(N_t-3)!} +..
\frac{q_1^{-(z^{*}-2)}\alpha(z^{*},L)\left(\frac{K_0}{\Lambda^3}\right)^{(z^{*}-2)} } {(N_t-z^{*})!}\right]\\
&=q_2^0 q_1^{N_t-2} \times \left[ \frac{\alpha(2,L)}{(N_t-2)!} + \frac{\alpha(3,L) \left(\frac{K_0}{V}\right)}{(N_t-3)!} +..
\frac{\alpha(z^{*},L)\left(\frac{K_0}{V}\right)^{(z^{*}-2)} } {(N_t-z^{*})!}\right]\\
&=\frac{q_2^0 q_1^{N_t-2}}{(N_t-2)!} \times \left[ \alpha(2,L) + \alpha(3,L) \left(\frac{K_0}{V}\right)(N_t-2) +..
\alpha(z^{*},L)\left(\frac{K_0}{V}\right)^{(z^{*}-2)}  \frac{(N_t-2)!}{(N_t-z^{*})!} \right]
\end{align}
and the partition function can further be transformed as
\begin{align}
Q^{\rm R}&=\frac{q_2^0 V^2}{K_0^2}\frac{q_1^{N_t-2}}{(N_t-2)!} 
\times \left[\alpha(2,L) \left(\frac{K_0}{V}\right)^2 +\alpha(3,L) \left(\frac{K_0}{V}\right)^3(N_t-2) +..
\alpha(z^{*},L)\left(\frac{K_0}{V}\right)^{z^{*}} \frac{(N_t-2)!}{(N_t-z^{*})!}\right]\\
&=\frac{q_2^0 V^2}{K_0^2}\frac{q_1^{N_t-2}}{N_t!} 
\times \left[\alpha(2,L) \left(\frac{K_0}{V}\right)^2 \frac{N_t!}{(N_t-2)!}+ \alpha(3,L) \left(\frac{K_0}{V}\right)^3 \frac{N_t!}{(N_t-3)!} +..
\alpha(z^{*},L)\left(\frac{K_0}{V}\right)^{z^{*}} \frac{N_t!}{(N_t-z^{*})!}\right]
\label{eq:rci2}
\end{align}

In the thermodynamic limit (T.L.), here $N_t \rightarrow \infty$, $A \rightarrow \infty$ with fixed ratio $N_t/A=\rho_t L$, one gets
\begin{align}
Q^{\rm R}(A,L,T,N_t,N_f=1)&=\frac{q_2^0 \Lambda^6}{K_0^2}\frac{q_1^{N_t}}{N_t!}  \times D(\hat{\rho}_t,L)
\label{eq:rci3}
\end{align}
where we have defined
\begin{align}
D(\hat{\rho}_t,L)&=\alpha(2,L)\hat{\rho}_t^2+\alpha(3,L)\hat{\rho}_t^3...+\alpha(z+1,L)\hat{\rho}_t^{(z+1)}+ ..+\alpha(z^{*},L)\hat{\rho}_t^{(z^{*})}\label{eq:dd}\\
\hat{\rho}_t&=\rho_t K_0=\frac{N_t K_0}{V}
\label{eq:rhot}
\end{align}
and where we have replaced $\frac{N_t!}{N_t^{i}(N_t-i)!}\approx 1$ in all terms of $D(\hat{\rho}_t,L)$.

We have thus
\begin{align}
\beta F^{\rm R}(A,L,T,N_t,N_f=1)&=N_t\;[\ln{\left(\Lambda^3 \rho_t\right)}-1]-\ln{\left(\frac{q_2^0 \Lambda^6}{K_0^2}\right)}-\ln{D(\hat{\rho}_t,L)}
\label{eq:rci4}
\end{align}

Note that in the T.L. we may replace $N_t$ by $N_1=N_t-l_{fil}$ (the average length of the filament) and $\rho_t$ by $\rho_1$, so that the first term in the r.h.s. of Eq.(\ref{eq:rci4}) is the Helmholtz free energy of the bath of free monomers. While the third term of Eq.(\ref{eq:rci4}) is the relevant free energy of the living filament, the middle term, function of $T$ only, must be linked to the free energy required to graft the filament seed (fixed dimer).

The probability for the living filament to have a size $j$, defined as $P(j) \equiv P(j;L,\hat{\rho}_t)$, is the term of index $j$ in the global partition function Eq.(\ref{eq:rci1}), properly normalized. Using the equivalent version of Eq.(\ref{eq:rci3}), leads to
\begin{align}
P(j)&= \frac{\alpha(j,L)\hat{\rho}_t^{j}}{D}
\label{eq:P}
\end{align}
for $j \in[2,z^{*}]$, where $D$ is given by Eq.(\ref{eq:dd}).

Performing a Legendre transform of the reactive Helmholtz free energy $F^R$ to the reactive grand potential $\Omega^{R}=F^{R}- \mu_1 N_t$, Eq.(\ref{eq:F}) becomes
\begin{equation}
d\Omega^{R}=-S dT - p_N A dL + p_T L dA - N_t d\mu_1 + (\mu_2-2 \mu1)dN_f
\label{eq:Ome}
\end{equation}
To obtain $\Omega^{R}$, one needs to express $\mu_1$ in terms of the old variables according to Eq.(\ref{eq:F}) using Eq.(\ref{eq:rci4}) for $F^{R}$ and associated $D$ and $\hat{\rho}_t$ Eqs. (\ref{eq:dd},\ref{eq:rhot}). We get
\begin{align}
\beta \mu_1&=\frac{\partial \beta F^{\rm R}}{\partial N_t}=\ln{\left(\Lambda^3 \rho_t\right)}-\frac{\partial{D(\hat{\rho}_t)/\partial \hat{\rho}_t}}{D(\hat{\rho}_t)} K_0/V \\
&=\ln{\left(\Lambda^3 \rho_t\right)}-\frac{\hat{\rho}_t\partial{D(\hat{\rho}_t)/\partial \hat{\rho}_t}}{D(\hat{\rho}_t)} \frac{1}{N_t}=\ln{\left(\Lambda^3 \rho_t\right)}-\frac{l_{fil}}{N_t}
\label{eq:mu1}
\end{align}
where we have introduced the average length of the filament
\begin{align}
l_{fil}(L,\hat{\rho}_t)=\frac{\sum_{j=2}^{z^{*}} j \;\alpha(j,L)\; \hat{\rho}_t^{j}}{D}
\label{eq:avlen0}
\end{align}

Formally the Legendre transform requires the inversion of Eq.(\ref{eq:mu1}) as $N_t=N_t(\mu_1)$ to estimate
\begin{align}
\beta \Omega^{\rm R}(A,L,T,\mu_1,N_f=1)&=\left[\beta F^{\rm R}(A,L,T,N_t,N_f=1) - N_t \beta \mu_1\right]_{N_t=N_t(\mu_1)} 
\label{eq:omega1}
\end{align}

Using Eqs.(\ref{eq:rci4},\ref{eq:mu1}), one gets successively

\begin{align}
\beta \Omega^{\rm R}(A,L,T,\mu_1,N_f=1)&=\left[ -N_t+l_{fil}\;-\ln{\left(\frac{q_2^0 \Lambda^6}{K_0^2}\right)}-\ln{D(\hat{\rho}_t)}\right]_{N_t=N_t(\mu_1)}\nonumber \\
&=\left[ -N_t+l_{fil}- \frac{l_{fil}^2}{N_t} \;-\ln{\left(\frac{q_2^0 \Lambda^6}{K_0^2}\right)}-\ln{D(\hat{\rho}_1)} \right]_{N_t=N_t(\mu_1)}
\label{eq:omega2bis}
\end{align}

where we have developed $D(\rho_t)$ around $\rho_1$ up to first order. We also note that Eq.(\ref{eq:mu1}) can be rewritten as
\begin{align}
\beta \mu_1=\ln{\left(\Lambda^3 \rho_1\right)}+ O\left(\frac{l_{fil}}{N_t}\right)^2\approx \ln{\left(\Lambda^3 \hat{\rho_1}/K_0\right)}
\label{eq:mu1f}
\end{align}
where the central expression is the expected relationship for the chemical potential of one species in an ideal mixture, the negligible correction coming from the approximations made earlier to simplify the $Q^R$ Eq. (\ref{eq:rci2}). 

Neglecting the term $\frac{l_{fil}^2}{N_t}$ in Eq.(\ref{eq:omega2bis}), the grand potential can finally be reformulated as
\begin{align}
\beta \Omega^{\rm R}(A,L,T,\mu_1,N_f=1)&=\left[ -N_1\;-\ln{\left(\frac{q_2^0 \Lambda^6}{K_0^2}\right)}-\ln{D(\hat{\rho}_1)} \right]_{\hat{\rho}_1=\frac{K_0}{\Lambda^3}\exp{(\beta \mu_1)}}
\label{eq:omega2ter}
\end{align}
In the biophysics literature it is customary to use the reduced free monomer density $\hat{\rho}_1$ as the independent variable instead of the more appropriate chemical potential $\mu_1$. Therefore we will re-express the grand potential in Eq.(\ref{eq:omega2ter}) as
\begin{align}
\beta \Omega^{\rm R}(A,L,T,\mu_1,N_f=1)
&= \beta \Omega^{\rm free}(A,L,T,\hat{\rho}_1) \ + \beta \Omega^{\rm fil}(L,T,\hat{\rho}_1)\label{eq:omegatot}\\
\beta \Omega^{\rm free}(A,L,T,\hat{\rho}_1)&=-\frac{AL}{K_0} \hat{\rho}_1\\
\Omega^{\rm fil}(L,T,\hat{\rho}_1)&=-k_BT\left[\ln{\left(\frac{q_2^0 \Lambda^6}{K_0^2}\right)}+\ln{D(\hat{\rho}_1)}\right]
\label{eq:omega2}
\end{align}
where we indentify the grand canonical contributions $\Omega^{\rm free}(A,L,T,\hat{\rho}_1)$ and $\Omega^{\rm fil}(L,T,\hat{\rho}_1)$ for the free monomers and the grafted living filament respectively.

The filament size distribution, the normalization factor $D$ and the average size of the filament, given respectively by Eqs.(\ref{eq:P},\ref{eq:dd},\ref{eq:avlen0}), take the final form
\begin{align}
P(j)&\equiv P(j;L,\mu_1)\equiv \frac{N_j}{N_f}=\frac{\alpha(j,L)\hat{\rho}_1^{j}}{D}
\label{eq:P2}\\
D(\mu_1)&= \sum_{j=2}^{z^{*}} \alpha(j,L)\hat{\rho}_1^{j}
\label{eq:dd2}\\
\bar{l}_{fil}(L,\mu_1)&=\frac{\sum_{j=2}^{z^{*}} j \;\alpha(j,L)\; \hat{\rho}_1^{j}}{D}= \hat{\rho}_1 \frac{\partial \ln D}{\partial \hat{\rho}_1}=\frac{\partial \ln D}{\partial \beta \mu_1}\label{eq:avlen}
\end{align}
where $\hat{\rho}_1$ in the r.h.s. is again used instead of $\mu_1$ and where $N_j$ is the average number of filaments with size $j$ within the  microscopic states of the reactive grand canonical ensemble.

\subsection{Single filament force exerted on the opposite wall}

Combining Eqs. (\ref{eq:Ome}) and (\ref{eq:omegatot}), and noting that $p_N A$ is the sum of a free monomer contribution and the single filament average force $f_{\bot}(L,\hat{\rho}_1)$, one gets
\begin{align}
\beta f_{\bot}(L,\hat{\rho}_1)&= -\left(\frac{\partial (\beta \Omega^{fil})}{\partial L}\right)= \left(\frac{\partial \ln{D}}{\partial L}\right)
 =\frac{\sum_{i=2}^{z^{*}} \frac{\partial \alpha(i,L)}{\partial L} 
(\hat{\rho}_1)^{i}}{D} \\
&= \sum_{i=2}^{z^{*}} \frac{\partial \ln{\alpha(i,L)}}{\partial L} 
P(i;L,\hat{\rho}_1) =\sum_{i=2}^{z^{*}} \beta \bar{f}_{i}(L) P(i;L,\hat{\rho}_1)\label{eq:fn}
\end{align}
where we have used Eq.(\ref{eq:P2}) and introduced a filament mean force potential
and associated mean force at fixed length 
\begin{align}
W_{i}(L)&=-k_{\rm B}T \ln{\alpha(i,L)}.
\label{eq:mp}\\
\bar{f}_{i}(L)&=
- \frac{\partial W_{i}(L)}{\partial L} 
\label{eq:mpforce}
\end{align}
Eq. (\ref{eq:fn}) gives the equilibrium force exerted on a living grafted
filament by a fixed planar wall located at a distance $L$ from the grafting wall. 
As expected, it is the average of the force exerted by the wall on a 
fixed length "dead" grafted filament (this latter force is an average over its internal degrees of freedom), weighted by the absolute probability $P(i;L,\hat{\rho}_1)$ of having a filament of length $i$. Of course, only the filaments sufficiently long to interact with the wall ($\alpha(i,L) \neq 1$) contribute to the average.

\subsection{Equilibrium constants and rates \label{reaction}}

Here, we discuss a few properties of the equilibrium constants valid for the arbitrary grafted flexible filament.

For the considered ideal mixture of $N_1$ free monomers and the series of $N_i$ grafted filaments of size $i$, one has\cite{b.Hill} 
\begin{align}
\beta \mu_1&=- \ln \left(\frac{q_1}{N_1} \right)= \ln{\Lambda^3}+\ln{\rho_1}\label{eq:mu1bis}\\
\beta \mu_i&=- \ln \left(\frac{q_i}{N_i} \right)= - \ln{\left(\frac{\alpha(i,L) q_i^0}{ N_f}\right)}+\ln{P(i)}\label{eq:mui}
\end{align}
using Eqs.(\ref{eq:q1},\ref{eq:alpha},\ref{eq:P2}).

Substituting Eqs.(\ref{eq:mu1bis},\ref{eq:mui}) in Eq.(\ref{eq:chemeq}) for any reaction given by Eq.(\ref{eq:reaction}), one gets 
\begin{align}
K_{i}(L,T) \equiv \frac{P(i)}{P(i-1)  \rho_1}=\frac{\alpha(i,L)}{\alpha(i-1,L)} \frac{q_i^0 \Lambda^3}{q_{i-1}^0}=\frac{\alpha(i,L)}{\alpha(i-1,L)} K_0(T)
\label{eq:k0i}
\end{align}
which defines the equilibrium constants $K_{i}$ and its link with the equilibrium constant $K_0$, already defined in Eq.(\ref{eq:K0}), which would apply in absence of wall. Eq.(\ref{eq:k0i}) expresses the evolution of the equilibrium constant $K_i(L,T)$ with increasing $i$, as a result of interferences between filaments of sizes $i$ and $(i-1)$ and the wall.

Some considerations on the related wall influence on the (de)polymerisation reaction rates are provided in appendix A%\ref{App_B}
, given their close connection to the equilibrium constants $K_i$. These rates become essential ingredients of the present approach when extended to the study of the coupling of a mobile wall dynamics and the filament (de)polymerization steps. 

\section{The Wormlike Chain model and the F-actin case.\label{actin}}

\subsection{The discrete model.}

To model the living grafted filament with fluctuating size in the range $2 \leq i \leq z^{*}$,  we adopt the d-WLC model with discrete contour length step $d$ and persistence length $\ell_p$. Using a cartesian reference frame where the grafting wall is at $x=0$ and the obstacle wall at $x=L$, the filament normally grafted at the wall at $x=0$ has its two first monomers located at $\bar{r_1}=(0,0,0),\bar{r_2}=(d,0,0)$. The filament with $i$ monomers, having a contour length $L_{c,i}=(i-1)d$ has a configuration fully specified by the set of coordinates $[\bar{r_j}]_{j=1,i}$ including the grafted dimer. The instantaneous internal potential energy of the filament of size $i$ is
\begin{equation}
E([\bar{r}]_i)= -(i-1)\epsilon_0' + \frac{\kappa}{d} \sum_{k=2}^{i-1}[1-\cos \theta_k]
\label{eq:energy}
\end{equation}
where $\kappa=k_B T \ell_p$ is the bending modulus of the filament, $\epsilon_0'$ the bonding energy associated to the chemical step Eq.(\ref{eq:reaction})\cite{RRMP.13} and $\theta_k$ the angle between 
successive bonds implying monomers ($k-1,k,k+1$). The configuration having the minimum energy corresponds to the straight filament with all bending angles at zero. The monomer-wall potential is zero or infinite depending whether the articulation point (monomer) $j$ is in the slab space ($0 \leq x_j \leq L$) or lies inside the obstacle wall ($x_j>L$). If we represent by $U^w$ the global filament-wall interaction potential, being the sum of all monomer-wall potentials, according to Eq.(\ref{eq:alpha}) the factors $\alpha(i,L)$ become
\begin{align}
\alpha(i,L)=< \exp{-[\beta U^w}]>_{i,0}=< \prod_{j=2}^{i} \Theta(L-x_j)>_{i,0}
\label{eq:alphadWLC}
\end{align}
where $<...>_{i,0}$ denotes a canonical average with weight $\exp{\left(-\beta E([\bar{r}]_i)\right)}$ (Eq.(\ref{eq:energy})) of a grafted filament of size $i$ in absence of the obstacle wall, and $\Theta$ is the Heaviside function. For this model, of contour length $L_{ci}=(i-1)d$, we have $q_i(L)=q_{i0}$ and hence $\alpha(i,L)=1$, as long as $i \leq z$, where $z$ is the integer given by
\begin{align}
z=INT(L/d)+1
\label{eq:zz}
\end{align}

In the case of a living filament undergoing (de)polymerizing reactions and for short enough filaments ($i \leq z$), this WLC model leads to the following expression for the equilibrium constant, as defined in Eq.(\ref{eq:K0}), \cite{RRMP.13}
\begin{align}
K_0&= 2\pi \exp{(\beta \epsilon_0')}\; 
\frac{d^4}{\ell_p}\;\left[1-\exp{(-2\ell_p/d)}\right]\approx  2\pi \exp{(\beta \epsilon_0')}\; 
\frac{d^4}{\ell_p}.
\label{eq:dWLC}
\end{align}
in terms of the fundamental parameters $d,\ell_p,\epsilon_0'$ of the filament model. The equilibrium constant for filaments hitting the wall is given by Eq.(\ref{eq:k0i}) where the $\alpha$ factors are given by Eq.(\ref{eq:alphadWLC}).

\subsection{Explicit calculations for the F-actin case.}

\subsubsection{The relevant $L$ and $\hat{\rho}_1$ regime to probe single F-actin polymerization force.}

The essential ingredients to get static properties of grafted actin filaments are the wall factors $\alpha(i,L)$ for a grafted d-WLC hitting a hard wall ($z^{*}> i >z$), with $\ell_p=5370d$ and $d=2.7nm$. On this basis, all properties can be derived for any supercritical value of the reduced monomer density $\hat{\rho}_1>1$.

We first consider the relevant range of wall position $L$ and the range of reduced free monomer concentrations $\hat{\rho}_1$ for which the polymerization force is operative and of interest for a quantitative comparison with in vitro experiments\cite{KP.04,Dogterom.07,DCBB.14}.
According to Mogilner\cite{Mo.09}, to produce a working force, individual filaments should be longer than $L_c=70nm$ (about 25 monomers) to avoid being too rigid but should remain below $L_c=500 nm$  (about 185 monomers) to avoid what he refers to as \textit{buckling}. In Footer et al. experiments\cite{Dogterom.07}, the polymerization force was measured for non buckled filaments of length $200nm$ (about 70 monomers). In filopedia bundles\cite{Mogil.05}, parallel filaments are cross linked by fascin but free portions of filaments at the leading edge are supposed to be of the order of $20-200nm$. Finally in the recent experiment of D\'emoulin et al.\cite{DCBB.14} the bundle length studied to get the velocity load relationship is of the order of $100-400nm$ (about $40-150$ monomers per filament) (see supplemental information of ref. \cite{DCBB.14}). Further, it has to be noted that in vitro experiments probe the polymerization force in moderate supercritical conditions ($1<\hat{\rho}_1<3$) to avoid too rapid buckling and interferences with spontaneous nucleation of new filaments \cite{Dogterom.07,DCBB.14}.

In our illustrative section of the F-actin case, we will concentrate on the supercriticality regime by considering two values of the reduced density $\hat{\rho}_1=1.7$ and $\hat{\rho}_1=2.5$. We will be interested to the wall position regime $20d <L< 100 d$ where actin filaments are sufficiently long to avoid unphysical influence of minimum size filaments ($j=2$) but still sufficiently short to avoid escaping filaments, as it will be made more precise later.

\subsubsection{The compression law of a grafted (fixed size) filament.}

\begin{figure}[ht]
\begin{center}
\includegraphics[angle=0,scale=0.40]{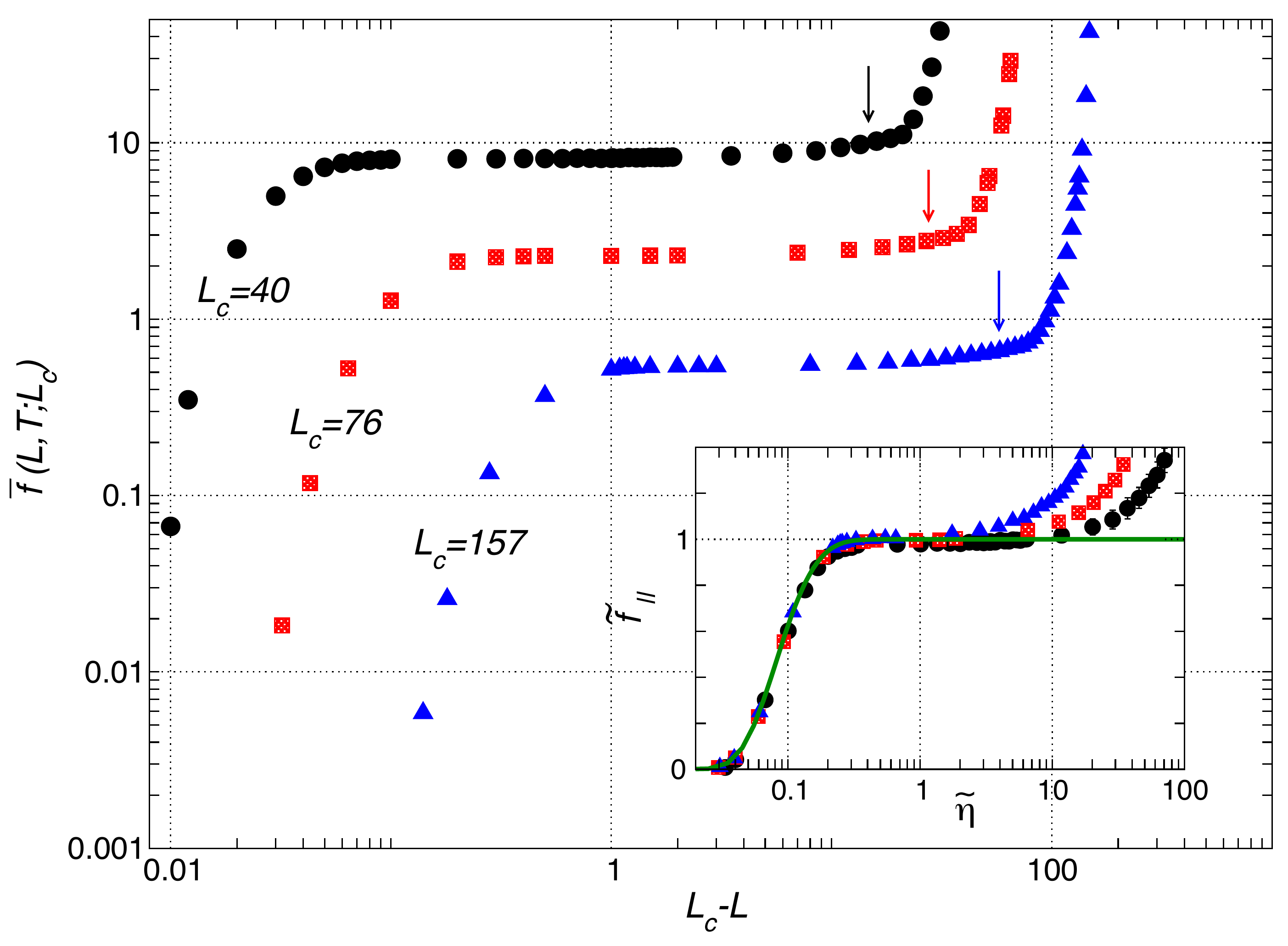}
\caption{Compressional force $\bar{f}(L,T;L_c)$ in units $k_BT/d$ exerted by a grafted dead d-WLC filament of contour length $L_c$ ($L_c=(n-1)d$ with $n=41, 77, 158$) and persistence length $\ell_p=5370d$ on an obstacle hard wall oriented normally to the filament grafting direction and located at distance $L$ from the filament's seed. For each filament size considered, we observe three successive regimes as $L$ decreases from $L_{c}$. First, a rapid rise corresponding to the weak bending regime, followed by a pseudo-plateau regime which terminates with the onset of the escaping filament regime characterized by a $1/L^2$ divergent behavior as $L$ approaches $0$ (as $(L_c-L)$ approaches $L_c$ on the Figure). The $L$ threshold value where the filament enters the escaping filament regime is indicated by a vertical arrow for each filament length (see text). In inset, all force data $\bar{f}(L,T;L_c)$ are reduced by the theoretical weak bending plateau value $f_{b}(L,L_c)$ for the continuous WLC model (see Eq.(\ref{eq:buc})) and plotted versus the renormalized compression distance $\eta$ (see Eq.(\ref{eq:etatilde})) in order to test the weak bending expression Eq.($\ref{eq:fpara}$) shown by a continuous green line.}
\label{elasticity}
\end{center}
\end{figure}

%\pdfmarkupcomment[markup=StrikeOut,color=red]{We consider the basic grafted filament compression law needed to feed our actin filament model with a realistic model for the $\alpha$ function defined by Eq.(\ref{eq:alpha}) for a fixed size filament of contour length $L_c=(n-1)d$ with arbitrary integer $n$.}{}
The basic input of our theory are the $\alpha$ functions (see Eq.(\ref{eq:alpha})) of "dead" filaments of contour length $L_c=(n-1)d$ ($z < n\leq z^*$).
The force $\bar{f}(L,T;L_c)$ exerted by a wall located at $L$ on a d-WLC filament with $\ell_p=5370d$ and of contour length $L_c \geq L$ (see Eq.(\ref{eq:mpforce})) (using here a notation without size index as we now deal with a unique dead filament hitting the wall) has been computed by Monte-Carlo simulation. The resulting force-compression laws for three filament sizes ($n=41, 77, 158$) are shown in Figure \ref{elasticity}. The MC sampling was realized by a mixture of two types of attempted moves, i) local crankshaft moves, where a sequence of three, four or five articulation points are rotated as a rigid body around an axis joining the two surrounding articulation points, and ii) pivot moves implying a global rigid rotation around a bond of the end chain fragment starting from that bond (the size of the fragment being sampled between $1$ and $(n-3)$ articulation points). The force exerted by the filament on the wall was estimated as 
\begin{align}
\bar{f}(L,T;L_c)/k_BT=\lim_{\Delta \rightarrow 0}\frac{1}{\Delta}\ln{\left[\frac{q(L,T;L_c)}{q(L-\Delta,T;L_c)}\right]}
\label{eq:f_mc}
\end{align}
where $q(L,T;L_c)$ is the partition function of a single grafted filament of contour length $L_c$. This force is easily estimated during the MC sampling by measuring the probability that the filament configuration has an articulation point located in the region of thickness $\Delta$ adjacent to the wall.

In Figure \ref{elasticity}, we observe that as $L$ decreases down from $L=L_c$ (where the force vanishes) the force quickly increases up to a pseudo-plateau before undergoing a final steep rise as $L$ approaches a value of $\approx 10d-15d$ on its way down to zero. As we are interested to filament lengths limited to $z^{*}=INT\left(\frac{\pi}{2} L/d\right)$ corresponding to the non escaping regime, it implies that only the elasticity law for the regime $L_c<L<\frac{2}{\pi}L_c$ needs to be exploited for each $L_c$. The relevant regime for our present study terminates within the pseudo plateau at $L_c-L=(1-2/\pi)L_c$ indicated by a vertical arrow for each of the lengths reported in the Figure. As it will be discussed elsewhere\cite{PPCR}, the fast increase of the elasticity force at short $L$ corresponds to a $L_c$ independent behavior $\bar{f}= A k_BT l_p/L^2$, with $A \approx 2/3$, valid for escaped filament lengths $L_c/d>>z^{*}$.

Gholami et al.\cite{Frey.06} derived a weak bending approximation for the elasticity law of a grafted continuous wormlike chain. Their prediction for the $L<<\ell_p$ regime, leads to the universal law summarized here. Using notations of reference\cite{Frey.06}, the identification with our formalism for a dead filament of size $n$ and contour length $L_c=(n-1)d$ leads to
\begin{align}
\alpha(n,L)&\equiv \tilde{Z_{\parallel}}(\tilde{\eta})
\label{eq:alpha_univ}
\end{align}
in terms of a new reduced compression distance 
\begin{align}
\tilde{\eta}=\frac{L_{c}-L}{L_{\parallel}}
\label{eq:etatilde}
\end{align}
involving the characteristic length 
\begin{align}
L_{\parallel}=\frac{L_{c}^2}{\ell_p}.
\label{eq:lpara}
\end{align}
The central quantity $\tilde{Z_{\parallel}}$ is (see eqs. (36) or (38) in reference \cite{Frey.06})
\begin{align}
\tilde{Z_{\parallel}}(\tilde{\eta})=2 \sum_{k=1}^{\infty}\left[(-1)^{k+1} \lambda_k^{-1} \exp{[- \lambda_k^2 \tilde{\eta}]}\right]
\label{eq:Z}
\end{align}
where $\lambda_k=(2k-1) \frac{\pi}{2}$. The (microscopically) averaged force $\bar{f}(L,T;L_c)$, defined by eq.(\ref{eq:mpforce}), which is the force exerted by the wall on a (dead) grafted WLC filament of contour length $L_{c}$ hitting the normal hard wall at seed-wall distance $L$, is\cite{Frey.06}
\begin{align}
\bar{f}(L;L_c)=f_{b}  \tilde{f}_{\parallel}(\tilde{\eta})
\label{eq:genfor}
\end{align}
where $f_{b}$ turns out to be equivalent to the buckling force for a clamped rod of contour length $L_{c}$ \cite{b.Howard}, namely
\begin{align}
f_{b}=\frac{\pi^2}{4} \frac{k_BT}{L_{\parallel}}
\label{eq:buc}
\end{align}
and where $\tilde{f}_{\parallel}(\tilde{\eta})$ is a universal function defined by 
\begin{align}
\tilde{f}_{\parallel}(\tilde{\eta})=-\frac{4}{\pi^2} \frac{\partial \ln{\tilde{Z_{\parallel}}(\tilde{\eta})}}{\partial \tilde{\eta}}
\label{eq:fpara}
\end{align}
This function, shown in the inset of Figure \ref{elasticity}, starts from $0$ at $\tilde{\eta}=0$ and increases monotonically to a unity plateau which is reached around $\tilde{\eta}=0.25$. We argue that the Gholami et al. elasticity function is worth exploiting not only for the weak bending regime (limited to $\tilde{\eta} \approx 0.6$) where it is rather precise, but also for the intermediate pseudo plateau regime up to $L_c/L=\pi/2 \approx z^{*}/z$ where the force appears to be underestimated by $10-15$ percent only. When this approximation is made for our purpose, the gain is enormous as we do not have to run a large number of single filament MC simulations to get the force for each specific filament size $n$ ($L_{c}=(n-1)d$) as a function of the continuous $L$ variable. We get all the needed expressions as functions of a single universal variable $\tilde{\eta}$ (Eq.(\ref{eq:etatilde})) under the form of an explicit convergent series easy to compute (see Eqs.(\ref{eq:Z},\ref{eq:fpara})).

\subsubsection{Living filament force in the grand canonical ensemble and the $L$ average force concept.}

Given the properties of the WLC discussed in the the previous subsection, the general expression of the polymerization force for the d-WLC model, Eq.(\ref{eq:fn}), takes the explicit form
\begin{align}
\beta f_{\bot}(L,\hat{\rho}_1)&=\sum_{i=z+1}^{z^{*}} \beta \bar{f}_{i}(L) P(i;L,\hat{\rho}_1)
\label{eq:fnwlc}
\end{align}
where $z$ is defined in Eq.(\ref{eq:zz}) and where $\bar{f}_{i}(L)$ and $P(i;L,\hat{\rho}_1)$ are given by Eqs.(\ref{eq:mpforce},\ref{eq:P2},\ref{eq:dd2}), computed with model functions Eqs.(\ref{eq:genfor},\ref{eq:alpha_univ}).

\begin{figure}[ht]
\begin{center}
\includegraphics[angle=0,scale=0.50]{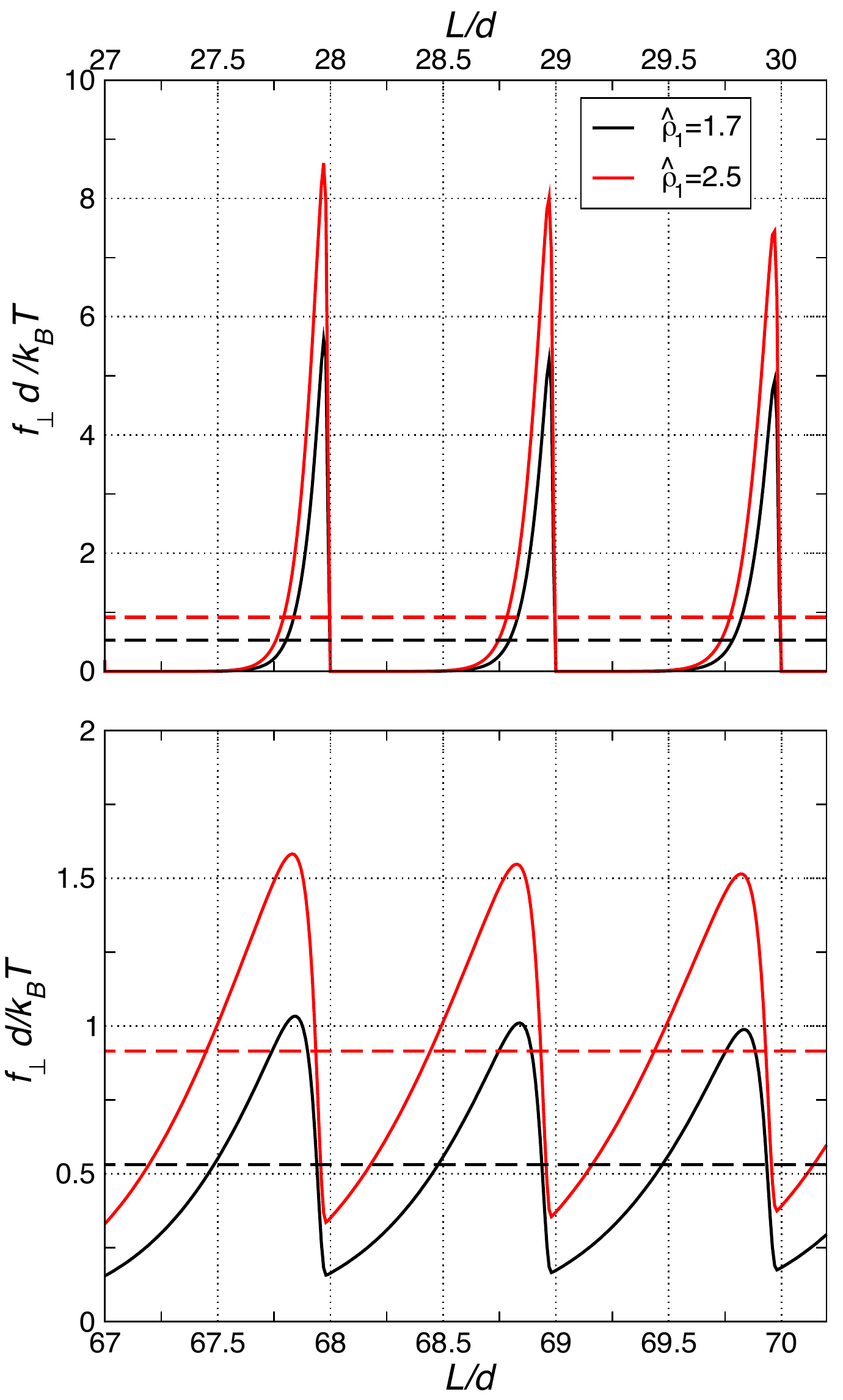}
\caption{$L$ dependence of the polymerization force Eq.(\ref{eq:fn}) for a F-actin single living filament, modeled by the living d-WLC. The force is shown for two values of the reduced free monomer concentration, $\hat{\rho}_1=1.7$ (black curve) and $\hat{\rho}_1=2.5$ (red curve), as a function of $L$ in the interval $[27d,30d]$ (upper panel) and in the interval $[67d,70d]$ (lower panel). The dashed horizontal lines show Hill's predicted value for the reduced force $\ln{\hat{\rho}_1}$ at each considered free monomer reduced concentration. %\JPR{ Carlo, can you indicate $f_{\bot}$ twice in ordinate.}
}
\label{fig:force_L}
\end{center}
\end{figure}

In Figure \ref{fig:force_L}, we report the polymerization force $f_{\bot}$ for the d-WLC model adapted to F-actin persistence length, as given by Eq.(\ref{eq:fnwlc}), for two values of $\hat{\rho}_1$ and highlighted within specific ranges of $L$. We observe a pseudo periodic signal of period $\lambda \approx d$. In the lower $L$ regime (upper panel), only the first term in the rhs of Eq.(\ref{eq:fnwlc}) ($i=z+1$) contributes to the force. To discuss the behavior of the polymerization force, let us focus on the interval where $L$ changes from $L=28d$ up to $L=29d$ (where $z=29$), noting that the force $f_{\bot}(L)$ is essentially zero at both boundaries. In this interval, $f_{\bot}(L) \propto \bar{f}_{30}(L) \alpha(30;L)/D(L)$. Given the moderate variation of $D$ with $L$ and the fact that the force $\bar{f}_{30}$ is essentially constant except when $L$ approaches the filament contour length $L_c=29d$ by less than $0.05d$ ($\tilde{\eta} \approx 0.3$) where it drops quickly to zero when $L$ reaches $L_c$, the rise of $f_{\bot}(L)$ reflects the $L$ dependence of $P(30;L,\hat{\rho}_1)$ and thus the dependence of $\alpha(30,L)$ from practically zero (for $L<28.5d$) towards unity. The fast drop of $f_{\bot}(L)$ towards zero at $L=29d$ comes from the drop of $\bar{f}_{30}$. As the $L$ domain increases (lower panel), the variation becomes more complex as it involves increasingly more terms in Eq.(\ref{eq:fnwlc}). We note in Figure \ref{fig:force_L} that Hill's prediction for the polymerisation force lies very close to the average value $f_{\bot}(L)$ within any interval.

In his seminal paper, Hill\cite{Hill.81} introduced the stalling force through the work needed to add reversibly a new monomer to a rigid filament pressing normally against a wall, as the wall moves by a distance $d$ (See Eq.2 in ref\cite{Hill.81}). This implicitly defines the average of the living filament force over an interval $[L,L+d]$. On the basis of a reversible change of the grand potential Eq.(\ref{eq:Ome}), the reversible work at constant $T$ and constant $\mu_1$ performed by the filament pressing against a wall moving from $L$ up to $L+d$ satisfies
\begin{align}
W_{L,L+d}(L,\mu_1,T)&=\Omega^{fil}(L,\mu_1,T)-\Omega^{fil}(L+d,\mu_1,T)=k_BT \ln{\left[ \frac{D(L+d,\hat{\rho}_1)}{D(L,\hat{\rho}_1)}\right]\equiv F_{fil}^{av} d}.
\label{eq:work}
\end{align}
where we have defined the average force $F_{fil}^{av}$ over the $d$ interval.

The average force concept appearing in Eq.(\ref{eq:work}) can be used for rigid filaments but also for flexible filaments as modeled by the d-WLC model, if one adapts specifically the $D(L,\hat{\rho}_1)$ terms (in particular the $\alpha$ functions) in Eq.(\ref{eq:dd2}). In order to do so, we first reexpress this average force in an equivalent but more appropriate way to discuss the specificities of filament flexibility. Given the definition of $z$ in Eq.(\ref{eq:zz}), we have $z(L+d)=z(L)+1$. For the upper limit $z^{*}$ given by Eq.(\ref{eq:z*}), one has $z^{*}(L+d)=z^{*}(L)+k$ where $k=1$ or $k=2$ depending upon the precise $L$ value which is truncated by the integer value operator in eq. (\ref{eq:z*}).
We can then rewrite $D(L+d)$ in alternative equivalent ways:
\begin{align}
D(L+d)&= \hat{\rho}_1 \left[\sum_{i=2}^{z(L+d)} \hat{\rho}_1^{i-1} + \sum_{j=z(L+d)+1}^{z^{*}(L+d)} \alpha(j,L+d)\  \hat{\rho}_1^{j-1}\right]\\
&= \hat{\rho}_1 \left[\hat{\rho}_1 + \sum_{m=2}^{z(L)} \hat{\rho}_1^{m} + \sum_{n=z(L)+1}^{z^{*}(L+d)-1} \alpha(n+1,L+d) \hat{\rho}_1^{n}\right]\\
&= \hat{\rho}_1 \left[\hat{\rho}_1 + \sum_{m=2}^{z(L)} \hat{\rho}_1^{m} + \sum_{n=z(L)+1}^{z^{*}(L)} \alpha(n+1,L+d)  \hat{\rho}_1^{n}+ \alpha(z^{*}(L+d),L+d) \hat{\rho}_1^{(z^{*}(L+d)-1)}\right]
\label{eq:interA}
\end{align}
Dropping the last term, present only when $z^{*}(L+d)-z^{*}(L)=2$ but which is anyway negligibly small in non escaping regime conditions, it leads to
\begin{align}
D(L+d)&\simeq \hat{\rho}_1 \left[\hat{\rho}_1 + \sum_{m=2}^{z(L)} \hat{\rho}_1^{m} + \sum_{n=z(L)+1}^{z^{*}(L)} \alpha(n+1,L+d)  \hat{\rho}_1^{n} \right]\label{eq:aven1} \\
\beta d F_{fil}^{av}(L,\mu_1,T) &=\ln{\hat{\rho_1}} ~+ \ln{\frac{[\hat{\rho_1}+D^{shift}(L)]}{D(L)}}
\label{eq:inter1}
\end{align}
where use of Eq.(\ref{eq:work}) has been made and where $D^{shift}$ is obtained by substitution of all terms $\alpha(j,L)$ in $D$ by $\alpha^{shift}(j,L)=\alpha(j+1,L+d)$ for all $j>z(L)$. Note that the first term in eq. (\ref{eq:inter1}) is the single filament polymerization force of Hill.

A similar calculation for the variation of the average size of the filament as the wall moves reversibly from $L$ to $L+d$ gives, according to Eqs.(\ref{eq:avlen},\ref{eq:work},\ref{eq:inter1}),
\begin{align}
\Delta \bar{l}_{fil}(L,\r1)\equiv \bar{l}_{fil}(L+d,\hat{\rho}_1)-\bar{l}_{fil}(L,\hat{\rho}_1)&=  \frac{\partial}{\partial \beta \mu_1}  \ln{\left[\frac{D(L+d,\hat{\rho}_1)}{D(L,\hat{\rho}_1)}\right]}=  \frac{\partial}{\partial \beta \mu_1} [\beta W_{L,L+d}]\label{eq:link} \\
&=1 + \hat{\rho}_1 \frac{\partial}{\partial \hat{\rho}_1} \ln{\left[\frac{[\hat{\rho}_1+D^{shift}(L,\hat{\rho}_1)]}{D(L,\hat{\rho}_1)}\right]}
\label{eq:lcorr}
\end{align}
where the first unity term is also the Hill's result for rigid filaments hitting normally the obstacle wall\cite{Hill.81}. 

In fact, in both Eqs.(\ref{eq:inter1}) and (\ref{eq:lcorr}), the second term on the r.h.s. gives the correction arising from two different effects. The first effect is the imposed minimal size of filaments which manifests itself at low $L$ by the additive term $\hat{\rho}_1$ in the numerator of the argument of the logarithm. The second effect linked to the ratio $\frac{D^{shift}(L,\hat{\rho}_1)}{D(L,\hat{\rho}_1)}$ is the effect of flexibility by opposition to the purely rigid case ($\ell_p=\infty$) where $D^{shift}=D$. Before embarking on this analysis, presented in Section \ref{comparison}, we derive the precise criteria to be satisfied, in supercritical conditions, to remain in the non escaping regime for the flexible filament case.

\subsubsection{Non-escaping regime criteria.\label{nonesc_crit}}

To avoid the presence of escaping filaments, one needs to have at equilibrium a negligible probability for filament size of the order of $z^{*}$. This can best stated by comparing this probability to the (near) maximum value at $i=z$ in supercritical conditions, giving
\begin{align}
\left[\frac{P(z^{*})}{P(z)}\right]=\alpha(z^{*},L) \hat{\rho}_1^{(z^{*}-z)}<<1
\label{eq:cond1}
\end{align}
where we have used Eq.(\ref{eq:P2}). Taking the logarithm of both sides and using Eq.(\ref{eq:mp}), one gets
\begin{align}
- \beta W_{z^{*}}(L) + (z^{*}-z) \ln{\hat{\rho}_1}  < 0
\label{eq:cond2}
\end{align}
The mean force potential $W_{z^{*}}(L)$ is the reversible work to compress a grafted filament of size $z^{*}$ until it fits within the space limited by a hard wall at $L$. Using the approximate universal expression of the force in the weak bending limit Eq.(\ref{eq:genfor}) and treating the plateau value $f_{bz^{*}}$ as constant over the whole compression interval, one gets 
\begin{align}
- \beta f_{bz^{*}}(\frac{\pi}{2}-1) L + (\frac{\pi}{2}-1) \frac{L}{d} \ln{\hat{\rho}_1} &<0,
\label{eq:inter}
\end{align}
Substituting Eq.(\ref{eq:buc}) for $f_{bz^{*}}$ in Eq.(\ref{eq:inter}), the final expression of the non escaping regime condition reads
\begin{align}
\hat{\rho}_1< \exp{\left(\frac{\ell_p d}{L^2}\right)}
\label{eq:escape}
\end{align}
which can be used either to limit $\hat{\rho}_1$ at given $L$ or to limit $L$ at given $\hat{\rho}_1$.

A comment about the evolution of size populations in the intermediate size window $z < i \leq z^{*}$ for any situation where condition (\ref{eq:cond1}) or equivalently (\ref{eq:escape}) is met  is in order.
According to Eqs.(\ref{eq:cond1}), (\ref{eq:cond2}), one has
\begin{align}
P(i)/P(z) \approx \exp{\left((i-z)\left[\ln{\hat{\rho}_1-\frac{\pi^2 \ell_p} {4 d}\frac{1}{(i-1)^2}}\right]\right)}
\hskip 1cm z<i 
\label{eq:pi_pz}
\end{align}
where we have again assumed the compressive force to be constant over the whole compression interval ($W_i(L)=(L_c-L)f_{bi}=[(i-1)d-L]\pi^2\ell_p/[4(i-1)^2d]$) and we have assumed $L\simeq(z-1)d$ according to Eq.(\ref{eq:zz}). The argument of the exponential is the product of a positive term $(i-z)$ and the factor in square brackets where the first constant and positive term is dominated by the negative second term at the lowest $i=z+1$ values as condition (\ref{eq:escape}) is met. According to Eq.(\ref{eq:pi_pz}), $P(i)$ must diverge as $P(i) \propto \hat{\rho}_1^{i}$ when $i$ grows to infinity. Therefore, the ratio in Eq.(\ref{eq:pi_pz}) must pass through a minimum (lower than unity) at some size $i_{min}$. So if $z^{*} <i_{min}$ (low $\hat{\rho}_1$ value), the ratio $P(i)/P(z)$ decreases monotonously over the relevant physical regime limited to $z^{*}$, down to a small value required by Eq.(\ref{eq:escape}). Otherwise, if $i_{min}$ is located in the relevant $z<i<z^{*}$ regime (higher $\hat{\rho}_1$ value), the criteria (\ref{eq:cond1}) implies that the ratio $P(i_{min})/P_z$ must be even lower than $P(z^{*})/P(z)$ so that kinetically, small filaments growing against the wall will see their size limited at values below $i_{min}$.

\begin{figure}[ht]
\begin{center}
\includegraphics[angle=0,scale=0.40]{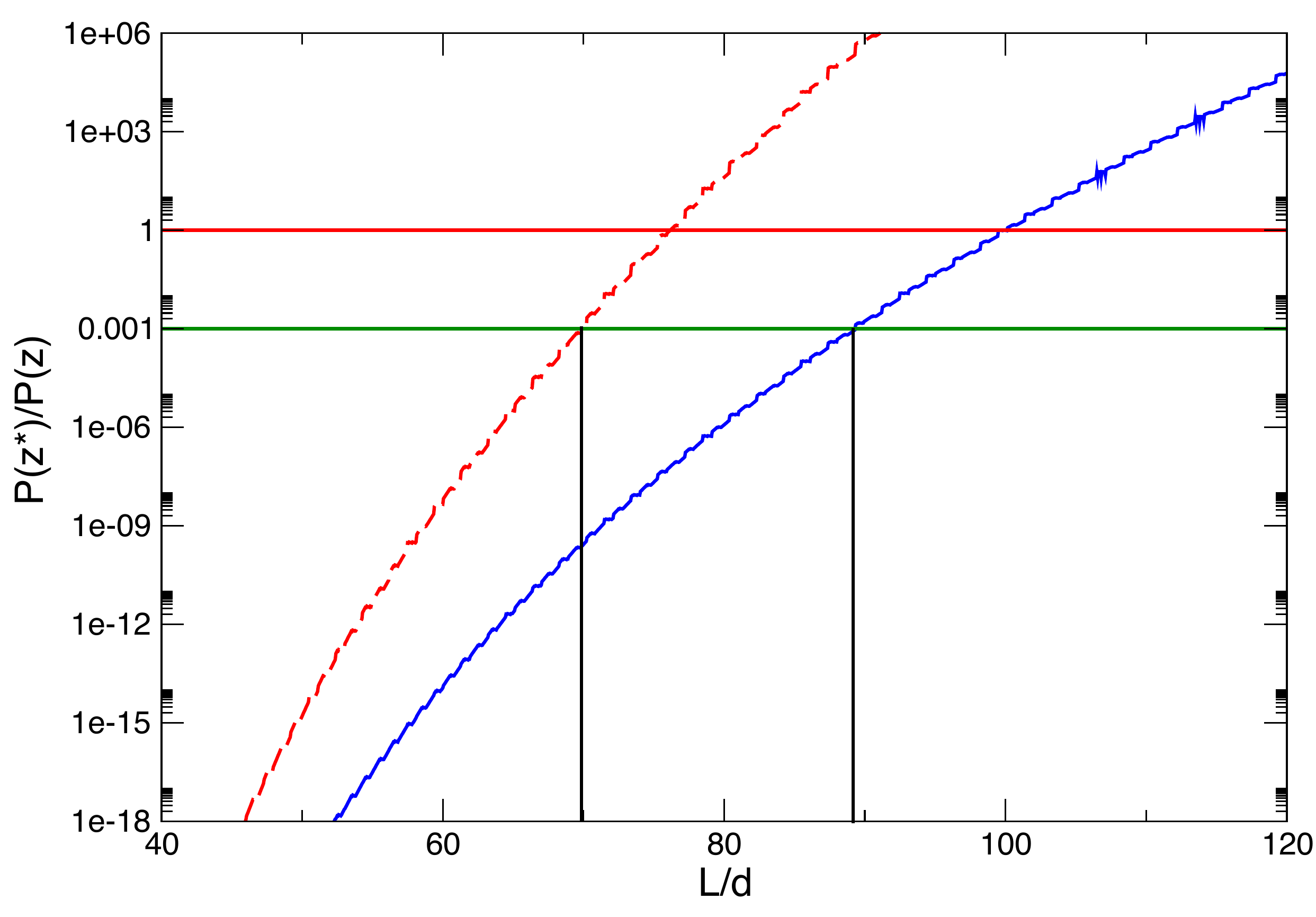}
\caption{Single F-actin filament with $\ell_p=5370d$. $L$ dependence of the ratio $P(z^*)/P(z)$ (in logarithmic scale) of the probability to have $z^*$ monomers over the probability to have $z$ monomers. We report the ratio for two values of the free monomer reduced density $\hat{\rho}_1=1.7$ (blue, continuous line) and $\hat{\rho}_1=2.5$ (red, dashed line). The horizontal green line marks the value $P(z^{*})/P(z)=0.001$ for which we consider that the bias induced by the constraint is negligibly small. The figure suggests maximal values of $L$ to be $L_{max}/d\simeq 89$ for $\hat{\rho}_1=1.7$ and $L_{max}/d\simeq 70$ for $\hat{\rho}_1=2.5$ (see also Eq.(\ref{eq:lmax})). The red horizontal corresponding to $P(z^{*})/P(z)=1$ level is reached for $L_l^2=\ell_p d/\ln{\hat{\rho}_1}$ which directly follows from Eq.(\ref{eq:escape}).}
\label{fig:pz}
\end{center}
\end{figure}

Eq.(\ref{eq:escape}) predicts that the range of $L$ values where the wall can effectively stop the bundle polymerization in supercritical conditions is limited by $L_l=\sqrt{\ell_p d/\ln{\hat{\rho}_1}}$, namely $L_l\simeq100$ and $L_l\simeq76$ for $\hat{\rho}_1=1.7$ and $2.5$ respectively. In practice, for fixed $\hat{\rho}_1>1$, we will limit the non escaping regime at the lower value $L_{max}$ which corresponds to $P(z^*)/P(z) = 0.001$, as illustrated in Figure \ref{fig:pz}. The dependence of $L_{max}$ upon $\r1$, empirically established, is
\begin{align}
L_{max}(\r1)&=L_l(\r1)-\Delta(\r1)=\sqrt{\ell_p d/\ln{\r1}}-\Delta(\r1)\label{eq:lmax}\\
\Delta(\r1)&=24.538-10.695~ \r1+ 1.3965~  \r1^2
\label{eq:De}
\end{align}
where the $\Delta$ term has been fitted in the $\r1$ range between $1.7$ and $4$. This relation provides $L_{max} \simeq 89$ and $L_{max} \simeq 70$ for $\hat{\rho}_1=1.7$ and $2.5$ respectively, as seen in Fig.\ref{fig:pz}.
As a point of comparison, the measurement of the polymerisation force in an optical trap set up for what appears to be a single actin filament in supercritical conditions at $\hat{\rho}_1=1.7$ \cite{Dogterom.07} (see Fig 4b of that reference), involves an elongation of $L \approx 200nm$ which corresponds to about $L=74d$ and is compatible with the present definition of the non-escaping regime.

Condition (\ref{eq:escape}) indicates that the concept of "non-escaping regime" is valid for flexible filaments only, since when $\ell_p\to \infty$ the inequality is satisfied for any finite $\hat{\rho}_1$ value.

%\section{Rigid versus Flexible filaments and Hill's expression of the stalling force for actin.\label{comparison}}
\section{The stalling force and energy conversion for F-actin.\label{comparison}}

\subsection{The rigid living filament case.}

\begin{figure}[]
\begin{center}
\includegraphics[scale=0.45]{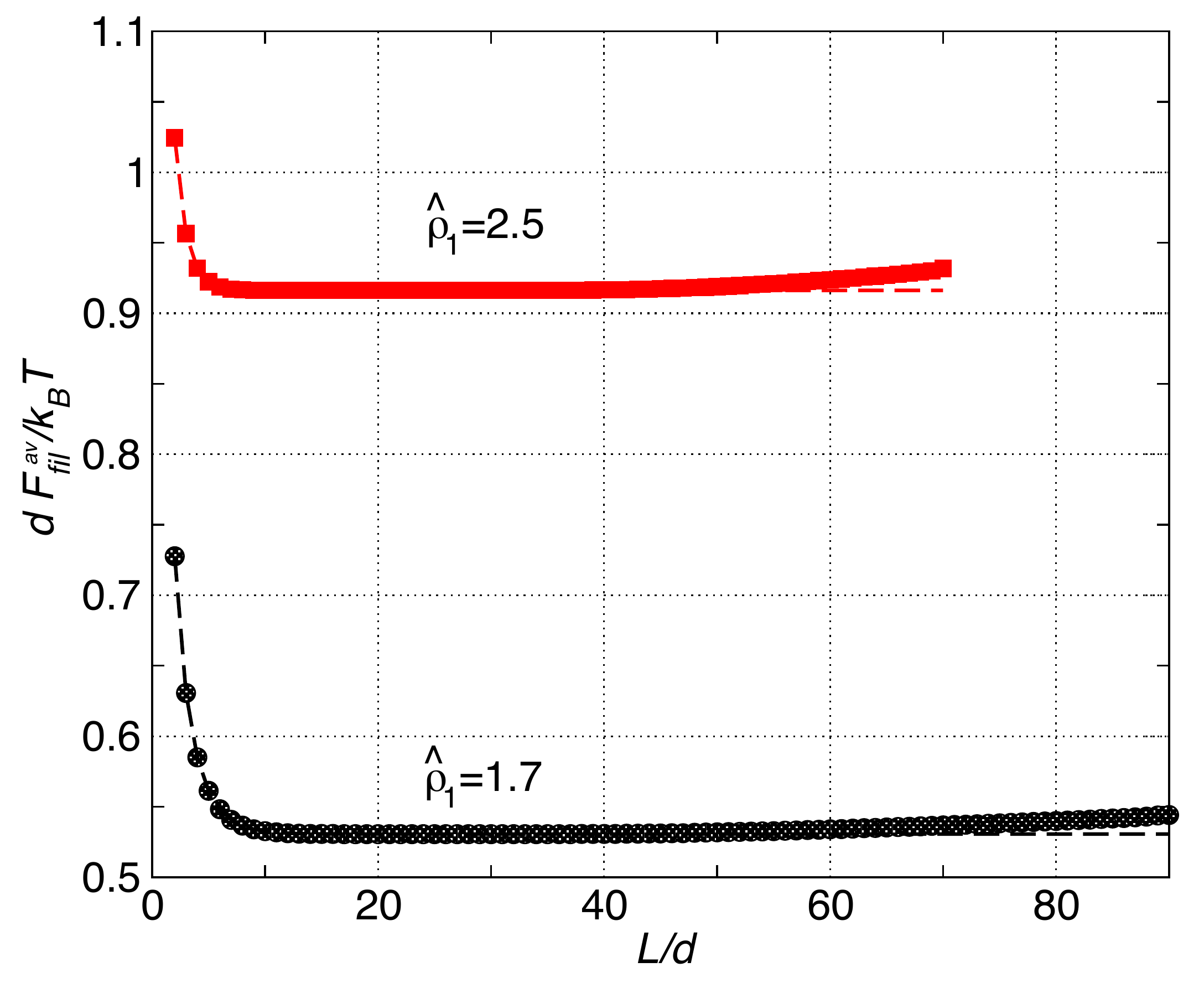}
\caption{The polymerization force, averaged over an interval from $L=nd$ to $L=(n+1)d$ and reduced by $k_BT/d$, is shown as a function of $n$ for a single living filament at supercritical conditions $\hat{\rho}_1=1.7$ (black curve) and $\hat{\rho}_1=2.5$ (red curve). At each density, the flexible d-WLC case ($\ell_p=5370d$, data points) is compared to the rigid limit case ($\ell_p=\infty$, dashed lines) on the basis of Eq.(\ref{eq:inter1}), which simplifies to Eq.(\ref{eq:avec}) for the rigid case. The peculiar short $L$ behavior, common to all curves, essentially reflects the boundary effect related to the imposition of a filament minimum size $i_{min}=2$. Beyond $L>\bar{L}$ (see text), the rigid filament average force goes to the asymptotic value $\ln{\hat{\rho}_1}$ in agreement with Hill's expression. Curves are deliberately interrupted in the Figure at $L=L_{max}$ which is the upper range of the non escaping regime according to Eq.(\ref{eq:lmax}).}
%\textit{fig:opt_trap}
\label{fig:rig3}
\end{center}
\end{figure}

The living filament polymerization force $ F_{fil}^{av}(L,\mu_1,T)$, shown for various cases in Figure \ref{fig:rig3}, is derived in Eq. (\ref{eq:inter1}) as the average of the $L$ dependent force over an interval equal to the single monomer size $d$. In the limiting case of a rigid filament, any $\alpha(i,L)$ is a step function being unity as long as $L \geq L_{ci}$ (implying $i \leq z(L)$) and zero otherwise. Hence one has, using the Eq.(\ref{eq:dd2}) of $D$  and Eq.(\ref{eq:inter1}) of $D^{shift}$,
\begin{align}
D^{shift}(L)= D(L)=\frac{\r1^2(1-\r1^{z-1})}{1-\r1}
\label{eq:dddrig}.
\end{align}
Using this relation, the average force in Eq. (\ref{eq:inter1}) can be recast for the rigid filament case as 
\begin{align}
\frac{d F_{fil}^{av}}{k_BT}=\ln{\hat{\rho}_1}+\ln{\left[1+\frac{\hat{\rho_1}-1}{\hat{\rho_1}^{z(L)}-\hat{\rho_1}}\right]}.
\label{eq:avec}
\end{align}
The correction (second) term in Eq.(\ref{eq:avec}) is numerically important at small $L$ only. For it to be of order $\epsilon$, one has to go beyond $\bar{L}$ given by 
\begin{align}
\bar{L}/d \approx INT\left(\frac{\ln(\r1-1)-\ln(\epsilon)}{\ln(\r1)}\right)
\label{eq:lbar}
\end{align}
which follows from Eq.(\ref{eq:avec}) and from the link between $z$ and $L$ in Eq.(\ref{eq:zz}). This boundary problem for rigid filaments is illustrated in Figure \ref{fig:rig3} where it can be observed that the Hill's result, $\ln{\hat{\rho}_1}$, is indeed valid asymptotically beyond a value of $L=\bar{L}\approx 8d$ computed from Eq.(\ref{eq:lbar}) at $\hat{\rho}_1=2.5$ for $\epsilon=0.001$.

\begin{figure}[]
\begin{center}
\includegraphics[scale=0.45]{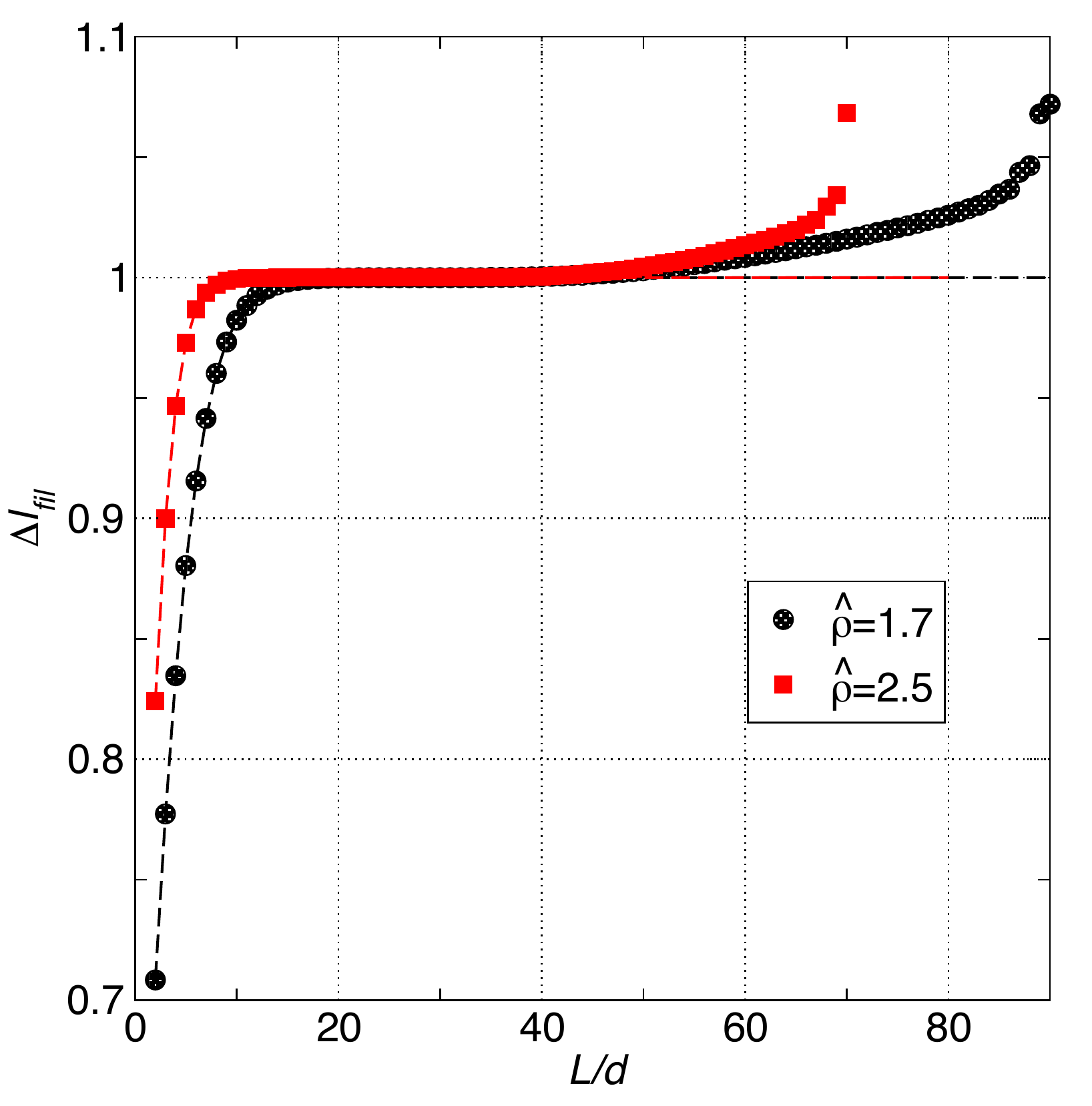}
\caption{
The increase $\Delta l_{fil}(L,\r1))$ of the average length of a living filament pressing against a wall as the latter is moved from $L=nd$ to $L=(n+1)d$ under supercritical conditions specified by $\hat{\rho}_1=1.7$ (black curve) or $\hat{\rho}_1=2.5$ (red curve) is reported according to Eq.(\ref{eq:link}). The rigid model case ($\ell_p=\infty$, dashed lines) (also given by Eq.(\ref{eq:lcorr1})) and the flexible case ($\ell_p=5370d$, data points) are compared. The behavior  at short $L$ ($L<\bar{L}$) results from the imposition of a lower end boundary condition on filament length, namely $i_{min}=2$.}
\label{fig:incr}
\end{center}
\end{figure}

\begin{figure}[]
\begin{center}
\includegraphics[scale=0.45]{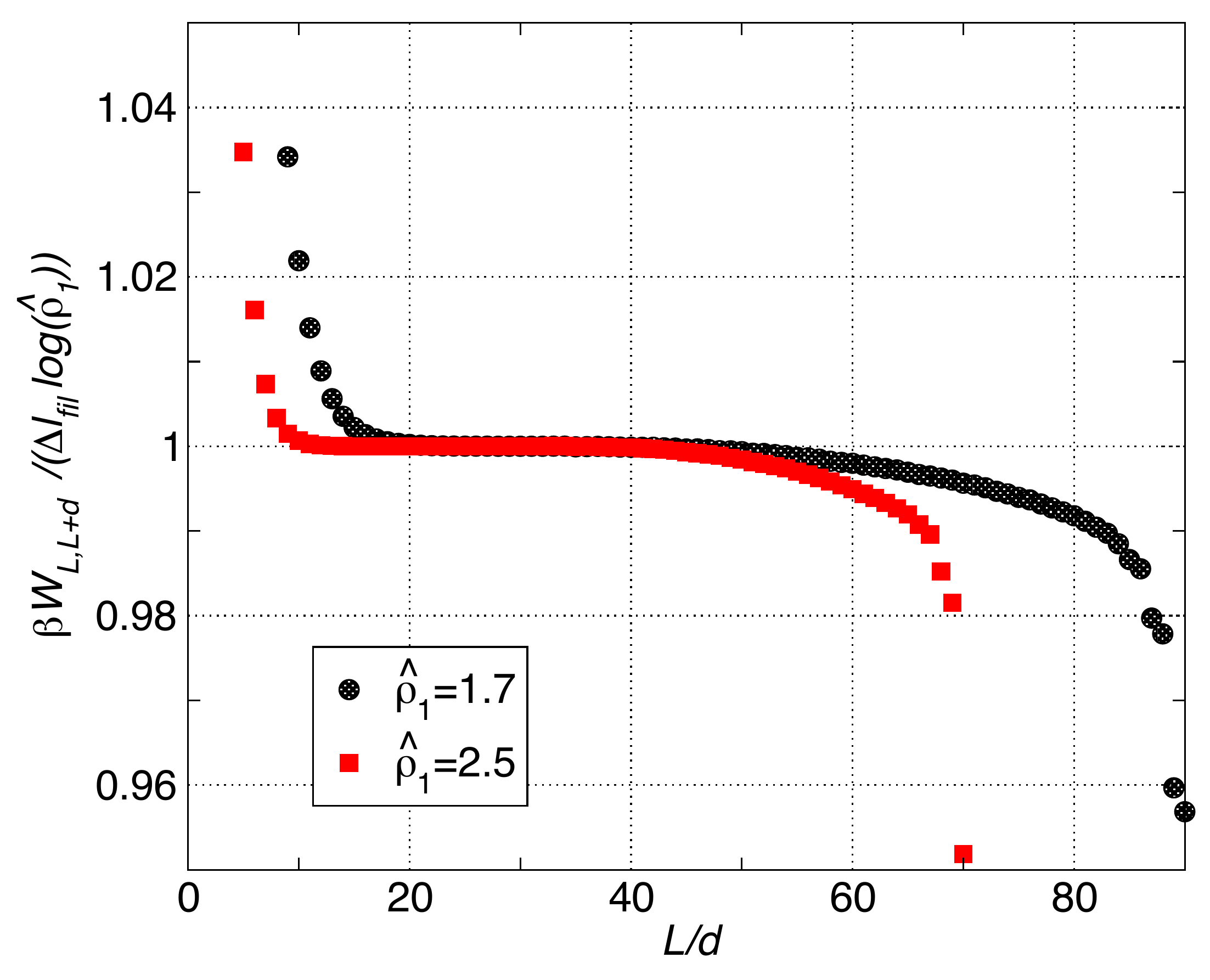}
\caption{$L$ dependence of the ratio of the reversible work of the polymerization force $W_{L,L+d}$ over the interval $L\in[nd,(n+1)d]$ and the corresponding chemical energy estimated as $ \Delta l_{fil}(L,\r1) k_BT \ln{\hat{\rho}_1}$. The results are shown for flexible filaments ($\ell_p=5370d$) at two free monomers reduced densities, $\hat{\rho}_1=1.7$ (black points) and $\hat{\rho}_1=2.5$ (red points). For $L>\bar{L}$, the observed central plateau value of unity reflects a perfect energy conversion. For the flexible filament case, the ratio starts decreasing progressively as $L$ approaches the upper limit of the non-escaping regime.}
\label{fig:ratio}
\end{center}
\end{figure}

Similarly in the rigid filament case, the size increment as the wall position $L$ is displaced by $d$ is given by combination of Eq.(\ref{eq:lcorr}) and  Eq.(\ref{eq:dddrig}), 
\begin{align}
\Delta l_{fil}(L,\r1))&=1 + \hat{\rho}_1 \frac{\partial}{\partial \hat{\rho}_1} \ln{\left[1+\frac{\hat{\rho_1}-1}{\hat{\rho_1}^{z(L)}-\hat{\rho_1}}\right]}\nonumber \\
&=1+\left[\frac{z}{\hat{\rho_1}^z-1}+\frac{(1-z)}{\hat{\rho_1}^{z-1}-1}\right]
\label{eq:lcorr1}
\end{align}
Again, the correction to unity vanishes for large $L/d$ ($L>\bar{L}$ where $\bar{L}$ is provided by Eq.(\ref{eq:lbar})). Figure \ref{fig:incr} shows the increment becoming asymptotically unity for the rigid filament case as $L$ increases. As commented by Hill\cite{Hill.81} and shown in Figure \ref{fig:ratio}, the ratio of the reversible work performed by the polymerization force to displace the wall by a distance $d$ over the corresponding chemical free energy $(\mu_1-\mu_{1c}) \Delta \ell_{fil}$ used to increase the average length of the filament, goes to unity asymptotically for the rigid filament case.


\subsection{The flexible living filament case adapted to F-actin.}

The effect of flexibility for a living grafted filament on the average polymerization force and on its size increment as the wall is displaced by the monomer size $d$ needs to be investigated for supercritical conditions in the regime $\bar{L}<L<L_{max}$. The higher limit, Eq.(\ref{eq:lmax}), was justified in section \ref{nonesc_crit} while the lower limit turns out to be in practice identical for flexible and rigid cases as illustrated in Figure \ref{fig:rig3} for two $\hat{\rho}_1$  values. Adopting $\epsilon=0.001$, one has for $\hat{\rho_1}=1.7$, $\bar{L}=12d$ and $L_{max}=89d$ and for $\hat{\rho_1}=2.5$, one has $\bar{L}=7.9$ and $L_{max}=70$.

Considering the general expression for the average force Eq.(\ref{eq:inter1}), we first establish that in the relevant $L$ regime and for the model of a WLC hitting a hard wall, the correction term is necessarily positive given the inequality
\begin{align}
D^{shift}(L)&> D(L)   \qquad(flexible \;filaments)
\label{eq:inter3}
\end{align}
This inequality basically follows from the property that $\alpha(i,L)$, as given by Eqs. (\ref{eq:alpha_univ}), (\ref{eq:Z}), is a monotonously decreasing function when $L$ decreases, or equivalently when $\tilde{\eta}$ increases (see Eq.(\ref{eq:etatilde})). This property is intuitively obvious and verified by visual inspection illustrative figures in ref. \cite{Frey.06}. To justify this on the basis of Eq.(\ref{eq:Z}), we note that it is a sum of decaying exponentials in terms of $\tilde{\eta}$ but with alternating sign. Grouping terms in pairs, the even $k-th$ term and the odd $(k+1)-th$ term, we obtain an absolutely converging series.
The justification of Eq.(\ref{eq:inter3}) follows from the property that, provided the sums of terms in $D(L)$ and $D^{shift}(L)$ converge sufficiently fast (before reaching $i=z^{*}$, thus well within the non-escaping regime), the two expressions can be compared term by term. The strict inequality $\alpha(j+1,L+d)>\alpha(j,L)$ for each pair of corresponding terms follows from the fact that, while $L_c-L$ is identical, the corresponding reduced compressions $\tilde{\eta}$ (Eq.(\ref{eq:etatilde})) is smaller for the $D^{shift}(L)$ term, implying a larger value for $\alpha$.
In Figure \ref{fig:rig3}, the average force $F_{fil}^{av}$ for a flexible actin living filament is shown as a function of $L$.
The resulting curve in the regime $\bar{L}<L<L_{max}$ is never very different from the rigid case, and thus from Hill's prediction.
%for large enough $L$. 
The averaged force is slightly above the Hill's plateau value by a marginal $1-2$ \% in the upper domain of the non escaping regime. This can be interpreted by noting that the $\alpha$'s are related to the fraction of possible chain conformations for a chain of given number of monomer in presence of a rigid obstacle at distance $L<L_c$. Intuitively this number should increase with the chain flexibility which is equivalent of taking longer chains $L_c+d$ at larger distance $L+d$ for given $\ell_p$.

The effect of flexibility on the filament size increment in the regime $L>\bar{L}$ gives, starting with Eq. (\ref{eq:lcorr}),
\begin{align}
\Delta \bar{l}_{fil}(L,\r1)&\approx 1 + \hat{\rho}_1 \frac{\partial}{\partial \hat{\rho}_1} \ln{\left[\frac{D^{shift}(L,\hat{\rho}_1)}{D(L,\hat{\rho}_1)}\right]}\nonumber \\
&=1+[\bar{l}^{shift}_{fil}(L,\hat{\rho}_1)- \bar{l}_{fil}(L,\hat{\rho}_1)]
\label{eq:lcorr2}
\end{align}
where $\bar{l}_{fil}(L,\hat{\rho}_1)$ is given by Eq.(\ref{eq:avlen}) and $\bar{l}_{fil}^{shift}(L,\hat{\rho}_1)$ is given by the same expression with probabilities $P(i)=\alpha(i,L)\hat{\rho}_1^{i}/D(L,\hat{\rho}_1)$ for $i>z$ substituted by $P(i)^{shift}= \alpha^{shift}(i,L)\hat{\rho}_1^{i}/D^{shift}(L,\hat{\rho}_1)$ (see also Eq.(\ref{eq:inter1})). 

On Figure \ref{fig:incr}, $\Delta \bar{l}_{fil}(L,\r1)$ computed with Eq.(\ref{eq:lcorr}) for the flexible case is again found to be close to the rigid limit. The value is however a few percents higher than unity in the upper part of the non escaping filaments domain, a logical result arising from the bending fluctuations of the filaments. Note that the approximate expression Eq.(\ref{eq:lcorr2}) (data not shown) gives identical results, except in the $L<\bar{L}$ domain.

In Figure \ref{fig:ratio}, the ratio of the reversible work of the polymerization force over the corresponding chemical free energy used to polymerize the living filament, goes also to unity for the flexible case at least in the central domain of the non escaping filament regime. At larger $L$, the ratio becomes lower than unity by a few percents, indicating that the conversion of chemical energy into work becomes affected by the flexible character of the filaments. Obviously, the situation quickly worsens if the filaments start to escape, which would happen with large probability if $L$ gets larger than $L_{max}$ by $5-10$ monomer units (Eqs.(\ref{eq:lmax},\ref{eq:De})).

\section{Distribution of filament sizes pressing against a fixed wall for F-actin. \label{distr}}

\begin{figure}[ht]
\begin{center}
\includegraphics[angle=0,scale=0.40]{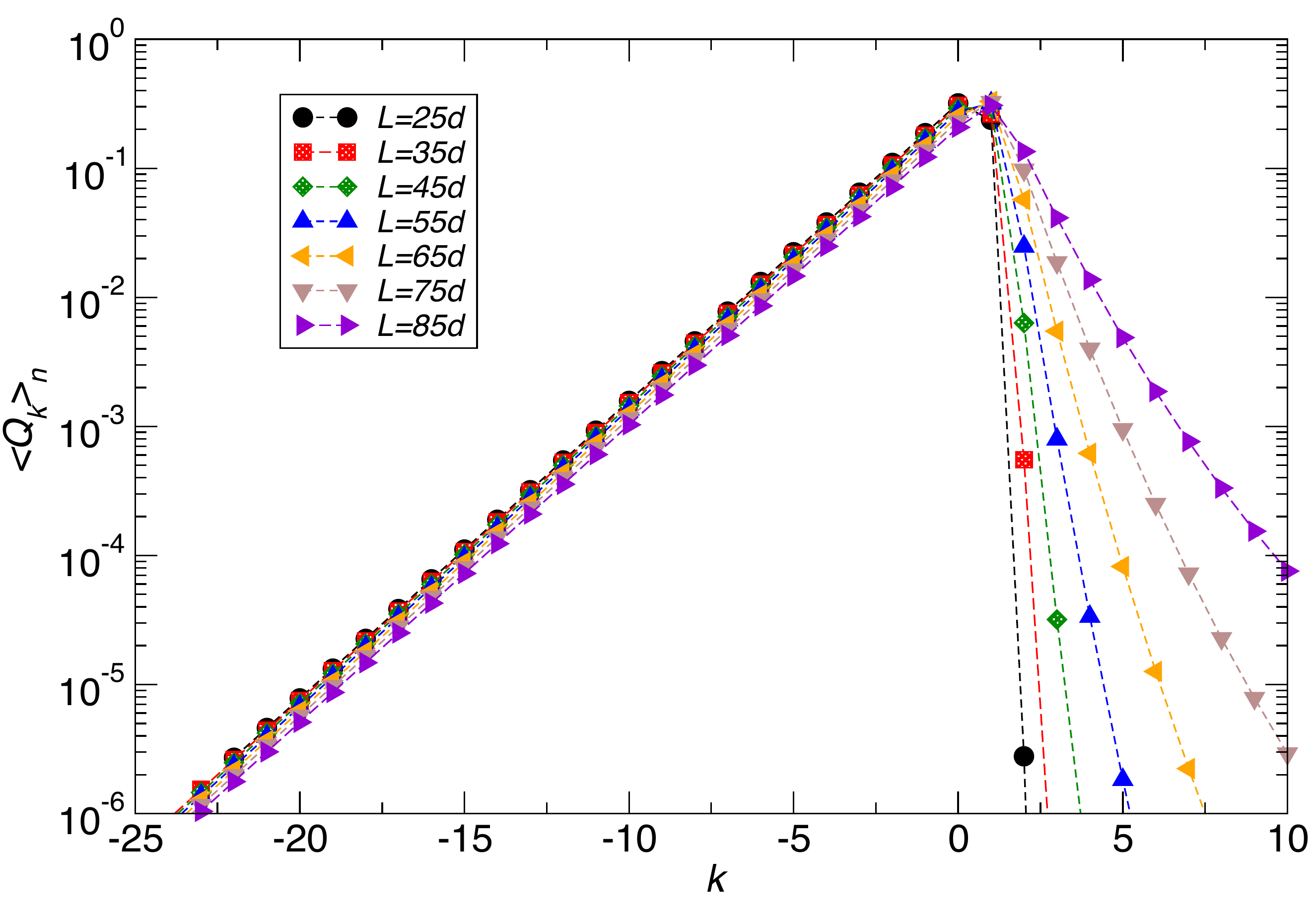}
\caption{Average normalized distribution $<Q_k>_n$ of filament relative sizes $k=i-z(L)$ at $\hat{\rho}_1=1.7$, for wall position averaged over the interval $[nd,(n+1)d]$ shown for various $n=L/d$ values within the non escaping regime. The exponential rise is observed for filaments sizes avoiding the wall, while the decay for filament sizes hitting the wall becomes increasingly sharper as $L$ decreases.}
\label{fig:lnpi}
\end{center}
\end{figure}

\begin{figure}[ht]
\begin{center}
\includegraphics[angle=0,scale=0.40]{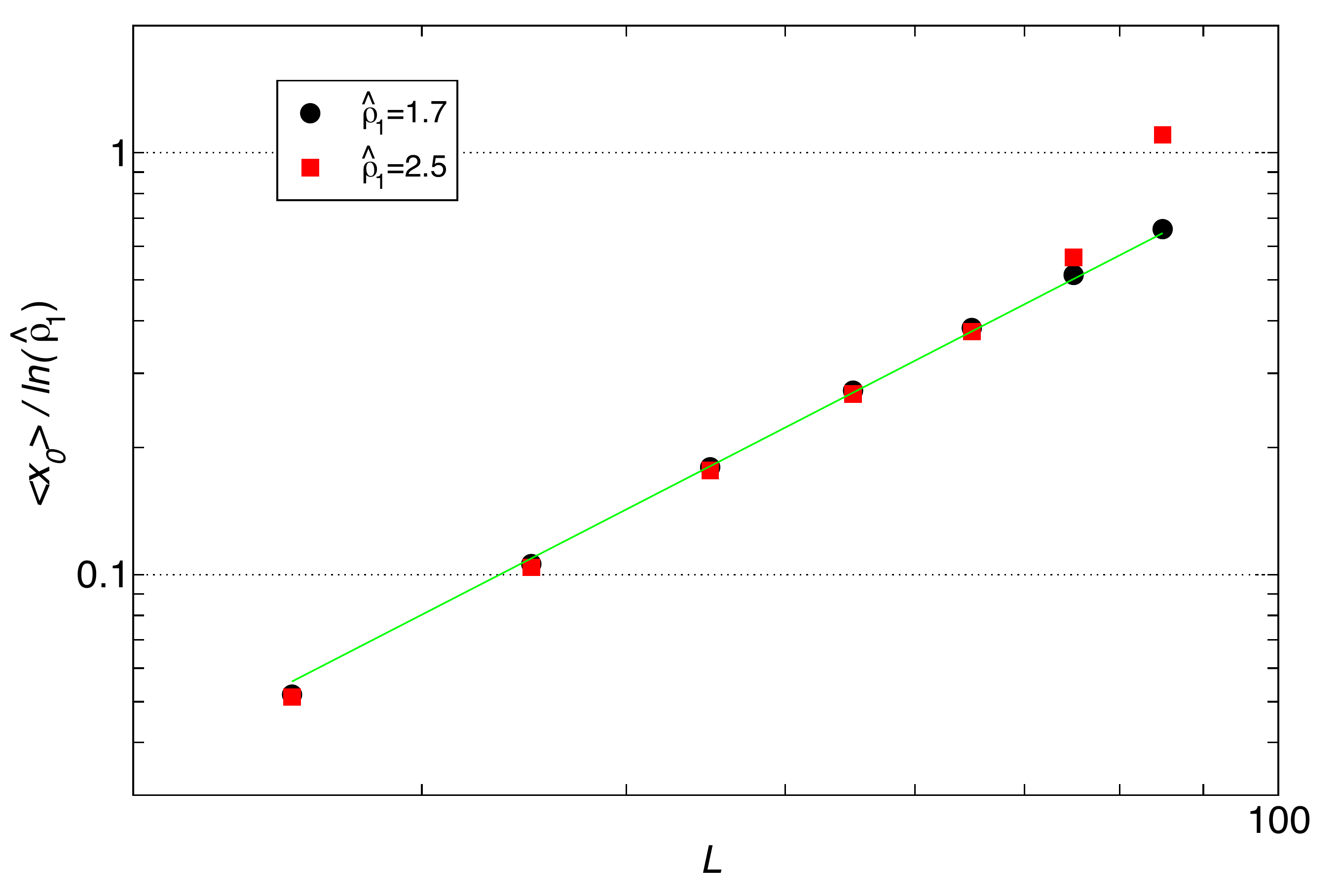}
\caption{Fraction $\langle x_0\rangle$ of filaments hitting the wall divided by $\ln{\hat{\rho}_1}$ plotted versus $L$ for two supercritical densities $\hat{\rho}_1=1.7$ (black circles) and at $\hat{\rho}_1=2.5$ (red squares). A clear trend $\langle x_0\rangle \propto \ln{\hat{\rho}_1} L^2$ for all data in the non escaping regime is indicated by the green continuous line.}
\label{fig:x0}
\end{center}
\end{figure}

In this last section, we analyze the influence of flexibility on the equilibrium distribution of filament sizes when a living filament in supercritical conditions, is stopped by a normal hard wall. The size distribution, given by Eqs.(\ref{eq:P2}), (\ref{eq:dd2}), takes in the rigid limit the form of a truncated growing exponential ($P(i;L,\hat{\rho}_1) \propto \exp{(i (\ln{\hat{\rho}_1}))}$ for $(i=2,z(L))$ and $P(i;L,\hat{\rho}_1)=0$ for $(i>z(L))$ and for $(i=1)$). 
In order to avoid large fluctuations of the probabilities of hitting filaments as $L$ varies over a monomer size distance $d$, we discuss results for the size distribution of flexible filaments in terms of an average over wall positions, as discussed earlier for the equilibrium polymerization force. Let $Q_k(L,\hat{\rho}_1)$ be the probability to have a filament of relative size $k=i-z(L)$ with respect to the fixed wall position $L$. The average $<Q_k>_n$, computed as the average of $Q_k$ over the interval $L/d\in[n,n+1]$, is shown in Fig. \ref{fig:lnpi} for several values of $L$ covering the entire non-escaping regime. In Fig. \ref{fig:x0} we show the average fraction $\langle x_0\rangle_n$ of filament sizes touching the wall, obtained as the cumulative sum of $<Q_k>_n$ over the positive values of $k$.

These results show the most spectacular features of semi-flexible filaments with respect to rigid case. The (average) equilibrium polymerization force $F_{fil}^{av}$ for filaments like F-actin in supercritical conditions and in the non escaping regime, is observed to remain essentially $L$ independent and equal to the standard rigid filament result of Hill $F_s^H=\frac{k_BT}{d}\ln{\hat{\rho}_1}$. However, the way this force is produced by the living filament is highly $L$ dependent and it is essentially obtained as the product of two factors with inverse $L$ dependencies
\begin{align}
F_{fil}^{av} \approx \langle x_0\rangle \frac{\pi^2}{4}\frac{k_BT \ell_p}{L^2}
\label{eq:central}
\end{align}
as seen in Fig. \ref{fig:x0}. This first order expression means that the required force is produced at wall position $L$ by recruiting a buckled filament of length $L_c \approx L$ with a weight $\langle x_0\rangle \propto \ln{\hat{\rho}_1} L^2$, while with weight $(1-\langle x_0\rangle)$, the filament does not contribute to the force on the wall, being it shorter than $L$. Indeed, filaments in contact with the wall are mostly in a compressed state corresponding to a force into the plateau region (see Fig. 2) therefore providing a force $f_{b}$ given by Eq.(\ref{eq:buc}) in the force expression Eq.(\ref{eq:genfor}). This means for a wall located at $L=50d$ that the plateau is reached as soon as $(L_c-L)>0.1d$ which is most often the case (see Fig. \ref{fig:lnpi}). In eq. (\ref{eq:central}) we also replaced $1/L_c^2 \approx 1/L^2$. Finally, it should be noticed that Eq. (\ref{eq:central}) also predicts that $\langle x_0\rangle \sim \ell_p^{-1}$ hence, in the rigid limit $\ell_p\rightarrow \infty$, $\langle x_0\rangle \to 0$ which demonstrates the impulsive character of the force when the rigid filament hits the wall during its Brownian fluctuations.

\section{Discussion and conclusions.\label{discussion}}

The ability of actin filaments to sustain in supercritical conditions a compressive force has predominantly been justified with the aid of the rigid living filament model, effectively a one dimensional model, both for the single filament case and for bundles of parallel filaments\cite{Hill.81,POO.93,MO.99,sander.00,Joanny.11,Krawczyk.11,Ca.01,DDP.14,DCBB.14}. The 1D filaments which are fluctuating in length as a result of (de)polymerizing steps are hitting a fluctuating obstacle usually subject to load by producing instantaneous kicks which result into a time averaged force biasing the obstacle brownian motion. These brownian ratchet dynamical models do lead to an effective polymerizing force, compatible with Hill's force expression at stalling, which satisfies a specific velocity-load relationship either for the single filament case \cite{POO.93} or for few bundle models differing mainly by the longitudinal disposition of filaments \cite{MO.99,sander.00,Joanny.11,DCBB.14}. When Hill's prediction for a bundle of $N_f>1$ actin filaments has been found to fail in interpreting experimental data \cite{Dogterom.07}, the role of flexibility has been invoked by introducing an ad-hoc maximum value for the force that a bunble can exert and beyond which the bundle buckles. However even this ad-hoc extension of the 1D model did not provide a satisfactory interpretation of the experiments.
%\pdfmarkupcomment[markup=StrikeOut,color=red]{The flexible character of living filaments like actin is then taken into account a posteriori by limiting the application of these rigid filament theoretical developments to situations where the average force remains below some estimated filament/bundle buckling threshold....}{}%\cite{Dogterom.07,DCBB.14}.

%\pdfmarkupcomment[markup=StrikeOut,color=red]{In the present paper and in its future extensions, we replace this ad-hoc combination of a rigid model treatment and some buckling limiting conditions by adopting from the beginning a flexible model}{} 
In the present paper and in its future extensions, we develop a Statistical Mechanics theory for a flexible filament model and we show that, quite generally, it leads to a systematic and continuous evolution of the filament behavior from a rigid rod character at short contour lengths to a pronounced flexible character at longer contour lengths, ending ultimately with the so called pushing catastrophe limit\cite{BM.05,BM.06}, when the filament(s) has(ve) acquired a finite probability to grow unimpeded by the wall and escape laterally. While our approach will involve ultimately multi-filament bundles, moving obstacles under various loads thus implying non equilibrium situations, in this work we have focused on the already very rich phenomenology offered by the basic equilibrium properties of a single grafted filament in supercritical conditions, as it hits a fixed wall oriented normally to its grafting direction. We have incorporated the living character and the flexibility of the filament explicitly into a statistical mechanics approach based on the reactive grand canonical ensemble, dealing with the model of a discrete WLC hitting a hard wall. This formalism has been illustrated by the F-actin/free G-actin reacting mixture restricted to a single kind of actin monomer-ATP complexes. 
%A few preliminary considerations on this model were already mentioned in previous work \cite{RRMP.13}.

The results and the new phenomenology which emerges from the present work can be summarized as follows
\begin{itemize} 
\item We provide a statistical mechanics justification of the popular Hill's expression for the single filament stalling force\cite{Hill.81}. Rigorously, this expression corresponds to the average force exerted by the filament (defined as the ratio of the mechanical work over a finite displacement and the displacement itself), as the wall moves reversibly, under mechanical and chemical equilibrium, over a distance corresponding to one monomer size $d$. The Hill's expression is found to be strictly valid only for the rigid filament case and we derive explicitly the correction terms for the semi-flexible case. The correction is positive and $L$-dependent for the model of a hard wall hit by a discrete WLC (the force is larger than for the rigid case) but these flexibility effects are only of the order of the percent when the experimental value of the actin persistence length is used. So we conclude that the $L$-independent Hill's expression remains a very good approximation for the stalling force of semi-flexible filaments like actin. It should be stressed however that the force exerted by the filament on a fixed wall in the reactive grand canonical ensemble, that is the force which was integrated to get the work over a finite displacement of size $d$, shows large fluctuations around the mean. These fluctuations which decrease progressively in amplitude as $L$ increases, find their origin in commensurability effects related to the degree of matching of the filament contour length, necessarily an integer number of monomer sizes $d$, with respect to the gap width $L$. These effects become less pronounced at large $L$ as the amplitude of tip transverse fluctuations due to bending become more important.
\item Like for the rigid case\cite{Joanny.11}, the equilibrium size distribution of the living flexible filament whose net polymerization is stopped in supercritical conditions by a wall at position $L$, starts as a growing exponential, as long as the filament size is too short for its set of fluctuating configurations to interact directly with the hard wall. Filament configurations larger than the slab gap have zero probability for the 1D rigid model, while for flexible filaments the size distribution generally presents a fast decay which involves some finite but rapidly decreasing probability to get filament contour lengths larger than $L$. This rapid decay results from a filament bending work penalty which systematically exceeds the gain in chemical free energy, as a result of polymerization steps beyond the largest size $z(L)$ of non touching filaments. Given the mentioned large oscillations in equilibrium properties as the gap width $L$ is varied over sub monomer length scales, the filament size distribution properties are better discussed in terms of their $d$-averaged (average over a $d$ window around the wall position $L$). The knowledge of these distributions (function of slab gap $L$) allows to adopt a quantitative definition for the limit of the non-escaping regime. In particular we require that the probability of a planar filament configuration of length $\pi L/2$, the minimum length to laterally escaping, be three orders of magnitude smaller than the probability of having filaments of length just below $L$.
%\pdfmarkupcomment[markup=StrikeOut,color=red]{The non escaping regime can then be precisely defined as the regime where the characteristic length of the final decay of the d-average size distribution, which systematically increases with $L$, remains well below the contour length extension $(\sqrt{\pi/2}-1)L$ between a straight filament of length $L$ and a filament coinciding with a quarter of a circle of radius of curvature $L$ needed for lateral escaping.}{}
Our work provides the opportunity to establish more precisely the characteristics of the crossover towards the escaping regime, a point of high relevance in in-vitro experiments\cite{Dogterom.07,DCBB.14}.

\item At stalling, in the non escaping regime, the quasi $L$ independence of the d-average force 
%(close to Hill's expression $F^H_{bun}$ for a single filament in Eq.(\ref{eq:force})) 
is produced by the fraction of filament configurations hitting the wall. The cumulative probability $\langle x_0\rangle(L,\hat{\rho}_1)$ of the size distribution involving hitting filaments has been shown to increase like $\langle x_0\rangle \propto \ln{\hat{\rho}_1}\; L^2/\ell_p$. This observation is compatible with the buckled filament state of the large majority of hitting filaments of the ensemble, each of them exerting adiabatically the classical plateau force expression $f_b=\frac{\pi^2}{4} k_BT \frac{\ell_p}{L_c^2}\approx  \frac{\pi^2}{4} k_BT \frac{\ell_p}{L^2}$ (here, adiabatic refers to the assumption that the life time of a given filament size is long with respect to the microscopic relaxation time of a fixed contour length filament). In this way the product $\langle x_0\rangle f_b$ is compatible with the $L$ independent stalling force expression of Hill, which allows us to pinpoint a major distinction between rigid and flexible living filaments, a distinction established here for the case of a single filament at equilibrium but which will be relevant for multi-filament bundles at and outside equilibrium. For finite $\ell_p$, the $L\rightarrow 0$ limit of very short semi-flexible filaments leads to a contact probability $\langle x_0\rangle\rightarrow 0$ and a buckling force $f_b \rightarrow \infty$, just like in the case of rigid filaments ($\ell_p = \infty$) at arbitrary $L$ . Hitting the obstacle takes the form of instantaneous kicks both for single filaments and multi-filament bundles. For flexible filaments of given $\ell_p$, the fraction of hitting filaments grows quadratically with $L$ while the force of each (buckled) filament decreases quadratically with $L$. 
If we consider a dynamical trajectory of a single living flexible filament at equilibrium against a wall at distance $L$, the fraction of time the filament is in contact with the wall is finite together with the associated (buckling) exerted force. 
For bundles of $N_f>1$ filaments at equilibrium, supposed to act independently, the force is produced by the permanent recruitment of a subset of $\langle x_0\rangle N_f$ filaments pressing each with the buckling force $f_b(L)$, the subset of hitting filaments permuting continuously among the $N_f$ equivalent filaments as the result of continuous (de)polymerization steps. Finally, as $L$ approaches $\sqrt{\ell_p d/\ln{\hat{\rho}_1}}$ under stalling conditions, which coincides with the upper limit of the non-escaping regime, the fraction $\langle x_0\rangle$ should approach unity as the polymerizing force exerted by a filament cannot exceed $f_b$. A more quantitative analysis of the limit is provided by eq. (\ref{eq:De}) which shows that  the probability to get escaping filaments starts to be non negligible when $\langle x_0\rangle$ approaches $0.5$. 
%\pdfmarkupcomment[markup=StrikeOut,color=red]{We have thus shown that, at stalling, this pushing catastrophe or escaping filament limit corresponds essentially to the crossover situation where the filament must press with its maximal force (the buckling force) permanently, instead of sporadically.}{}
\end{itemize}

A conjecture is possible when extending the criteria to observe the pushing catastrophe to the stationary situation of a wall moving at constant velocity, pushed by the polymerizing bundle of $N_f \geq 1$ filaments and subject to a load $F_L=\gamma F_{stal}$ smaller than the stalling value ($\gamma <1$) \cite{BM.05,BM.06}. To keep a constant velocity of the wall, the bundle must exert a force equal to the load $F_L$ (or sightly larger if solvent friction is considered), therefore the number of filaments $N_0=\langle x_0\rangle N_f$ needed to press on the wall should go as $N_0(L) \approx \gamma F_{stal}/f_b(L)\sim L^2$ in order to compensate for the variation of the single filament force $f_b\sim 1/L^2$. If we assume that the non-escaping limit in stationary conditions ($v>0$) would correspond to the recruitment of all $N_f$ filaments (or a permanent contact with the wall for the single filament case $N_f=1$), the maximum gap tolerated should be at least a factor $\gamma^{-1/2}$ larger than the limiting value for the non-escaping regime at stalling. 

It is illuminating to compare the above considerations with two reported experimental measurements of the actin polymerizing force. The experiments of Footer et al. \cite{Dogterom.07} use an optical trap set up to measure the stalling force of a few actin filaments in supercritical conditions. In particular, they report in Figure 4b data corresponding to a polymerizing force of $F\approx 1pN$ at reduced concentration $\hat{\rho}_1=1.7$ which, as they observe, corresponds to the stalling force of a single actin filament. The average filament length is $\approx 180 nm$ for the chosen optical trap. Eq.(\ref{eq:central}) applied to this case would imply a contact time fraction of $\langle x_0\rangle\approx 0.25$ with a force intensity of $f_b\approx 4pN$, the probed filament length being indeed lower than the limit $L_{max}=240 nm$ predicted by Eq.(\ref{eq:De}). This experiment was in fact dealing with a bundle of $N_f=8$ filaments but surprisingly enough the stalling force of a single filament was effectively recorded. The issue here is still under debate but, according to our present work, to detect a force eight times larger at the same reduced concentration in free monomers, a trap force constant $5-10$ larger would be required in order to avoid laterally escaping filaments. 

The Demoulin et al. experiment \cite{DCBB.14} probes the force-velocity relationship for a set of actin bundles\cite{DCBB.14}, implying a total of $N_f \approx 130$ filaments at $\hat{\rho}_1 \approx 3$, pressing together against a bead. While the stalling force in this case is around $200 pN$, the bead is subject to load forces ranging from a few $pN$ (largest velocity probed) up to $100 pN$ covering a range $0.02 <\gamma <0.5$ for the load over stalling forces ratio. Looking at figure 2 in ref.\cite{DCBB.14}, if we take a typical length of $200nm$ for the actin filaments beyond their lateral connection by fascin bridges, the number of filaments at contact able to press on the obstacle bead with a buckling force of $\approx 4pN$ should lie between $1$ and $25$ over the explored force range. Further, considering the results for the longest filaments $(\approx 400nm)$ at $F_L=3.9pN$, our criterium above for stationary non-escaping conditions is still justified since $L_{max}(\rho_1)=65d=175nm$ at $\rho_1=3$, and $\gamma^{-1/2}\approx 7$ in these conditions leading to a maximum length of the stationary non-escaping regime of $L_{max}/\gamma^{1/2}=1225nm$, still larger than the probed bundle length.

Further consequences of filament semi-flexible character on actin bundles at and outside equilibrium will be discussed in future publications\cite{PPCR,HMPCR}.

\section{Acknowledgements} 
We thank M. Baus, M. J. Footer and B. Mognetti for useful discussions and G. Destree and P. Pirotte for technical help. 
This work has been supported by the Italian Institute of Technology (IIT) under the SEED project Grant 259 SIMBEDD and by the Italian Ministery of Research under project PRIN2012  -- 2012NNRKAF.

\appendix
\section{Some considerations on (de)polymerization rates.\label{App_B}}

Phenomenological kinetic rate constants $k_{on}^{(i-1)}$ and $k_{off}^{i}$ are usually associated to the reactions described by Eqs.(\ref{eq:reaction}). In terms of such kinetic constants the equilibrium micro-reversibility conditions, i.e. the equality between the number of polymerizations of filaments of size $(i-1)$ to size $i$ and the number of de-polymerizations of filaments of size $i$ to size $(i-1)$ per unit of time, are written as
\begin{align}
k_{on}^{(i-1)} \rho_1 P(i-1) = k_{off}^{i} P(i)
\label{eq:kk}
\end{align}
which implies, according to Eq.(\ref{eq:k0i}) and thermodynamics, the link 
\begin{align}
K_i=k_{on}^{(i-1)}/k_{off}^{i}
\label{eq:ratiok}
\end{align}
Eq.(\ref{eq:kk}) is often written equivalently as 
\begin{align}
U_{(i-1)} P(i-1)  = W_i P(i)
\end{align}
where $U_{(i-1)}\equiv k_{on}^{(i-1)} \rho_1$ and $W_i\equiv k_{off}^{i}$ are (de-)polymerisation rates, respectively. Using again Eq.(\ref{eq:k0i}), their ratio is 
\begin{align}
\frac{U_{(i-1)}}{W_i}= \frac{P(i)}{P(i-1)} = \hat{\rho}_1 \frac{\alpha(i,L)}{\alpha(i-1,L)}
\label{eq:ratiow}
\end{align}
The (de)polymerization rates for filament ends in bulk, denoted by $U_0$ and $W_0$ with $U_0/W_0=\hat{\rho}_1$, are valid for our grafted filaments, as long as they do not interact with the wall. When $\hat{\rho}_1=1$ (that is the free monomer density is critical $\rho_1=\rho_{1c}=1/K_0$), one has $U_0=W_0$ so that the rate of polymerization and rate of depolymerization are equal and the distribution should be uniform in the short filaments region ($\alpha=1$). We are interested to the supercritical regime $\hat{\rho}_1>1$ and $U_0>W_0$ for which the distribution in the same short filament region is a growing exponential.
For (de)polymerization reactions implying filaments hitting the wall, the rates satisfying Eq.(\ref{eq:ratiow}) are often chosen in applications assuming that the rates of depolymerisation are not affected by the presence of the wall \cite{RR.13,POO.93,BM.06,Joanny.11}, namely
\begin{align}
W_i&=W_0 \\
U_{i-1}&=\frac{\alpha(i,L)}{\alpha(i-1,L)} U_0
\label{eq:rate_eff}
\end{align}
\newpage

%%%%%%%%%%%%%%%%%%%%%%%%%%%%%%%%%%%%%%%%%%%%%%%%%%%%%%%%%%%%%%%%%%%%%%%%%%%%%%%%
\nocite{*}
%\bibliography{refs}% Produces the bibliography via BibTeX.

%Merlin.mbs v4.21 2009-07-09.
%
%
%%%%%%%%%%%%%%%%%%%%%%%%%%%%%%%%%%%%%%%%%%%%%%%%%%%%%%%%%%%%%%%%%%%%%%%%%%%%%%%%
\end{document}